\documentclass[namedreferences]{SolarPhysics}
\usepackage[optionalrh,solaenum]{kd-sola-addons}
\usepackage{graphicx}
\usepackage{amssymb}
\usepackage{longtable}
\usepackage{url}
\newcommand{\Bl}{B_{\rm los}}
\newcommand{\Bt}{B_{\rm trans}}
\newcommand{\Bz}{B_{\rm z}}
\newcommand{\Br}{B_{\rm r}}
\newcommand{\Bh}{B_{\rm h}}
\newcommand{\Brp}{B_{\rm r}^{\rm pot}}
\newcommand{\Bzppc}{B_{\rm z}^{\rm pot,center}}
\newcommand{\Brppa}{B_{\rm r}^{\rm pot,all}}
\newcommand{\Brpsph}{B_{\rm r}^{\rm pot,sph}}
\newcommand{\Bzm}{B_{\rm z}^{\mu}}
\newcommand{\Brms}{B_{\rm r}^{\mu(s)}}
\newcommand{\grad}{\mbox{\boldmath$\nabla$}}

\newcommand{\BB}{\mbox{\boldmath$B$}}

\begin{document}

\begin{article}

\begin{opening}

\title{Evaluating (and Improving) Estimates of the Solar Radial Magnetic Field Component from Line-of-Sight Magnetograms}

\author{K.~D.~\surname{Leka}$^{1}$\sep
	G.~\surname{Barnes}$^{1}$\sep
	E.~L.~\surname{Wagner}$^{1}$\sep}
\runningauthor{Leka, Barnes, \& Wagner}
\runningtitle{Estimates of $\Br$ from $\Bl$}
\institute{$^{1}$ NWRA, 3380 Mitchell Ln., Boulder, CO 80301, USA
email: \url{leka@nwra.com} email: \url{graham@nwra.com}}

\begin{abstract}

Although for many solar physics problems the desirable or meaningful boundary
is the radial component of the magnetic field $\Br$, the most readily
available measurement is the component of the magnetic field along the
line-of-sight to the observer, $\Bl$.  As this component is only equal
to the radial component where the viewing angle is exactly zero, 
some approximation is required to
estimate $\Br$ at all other observed locations.  In this study, a common approximation
known as the ``$\mu$-correction'', which assumes all photospheric field to be 
radial, is compared to a method which invokes computing a potential field that matches
the observed $\Bl$, from which the potential field radial component, $\Brp$ is recovered.
We demonstrate that in regions that are truly dominated by radially-oriented
field at the resolution of the data employed, the $\mu$-correction
performs acceptably if not better than the potential-field approach.
However, it is also shown that for any solar structure which includes horizontal
fields, {\it i.e.} active regions, the potential-field method better
recovers both the strength of the radial
field and the location of magnetic neutral line.

\end{abstract}

\end{opening}

\section{Introduction}
\label{sec:intro}

Studies of the solar photospheric magnetic field are ideally performed
using the full magnetic field vector; for many scientific investigations
which may not require the full vector, it is the radial component $\Br$
which is often desired, as is derivable from vector observations.  Yet
observations of the full magnetic vector are significantly more difficult
to obtain, from both instrumentation and data-reduction/analysis points
of view, than obtaining maps of solely the line-of-sight component of
the magnetic field $\Bl$.  Observations consisting of the line-of-sight
component can accurately approximate the radial component only along
the Sun-Earth line, that is where the observing angle $\theta=0$, or
$\mu=\cos(\theta)=1.0$.  Away from that line, {\it i.e.} at any non-zero
observing angle, the line-of-sight component deviates from the radial
component.  Observing solely the line-of-sight component of the solar
photospheric magnetic field implies that the observing angle $\theta$
imposes an additional difficulty in interpreting the observations --
it is not simply that the full strength and direction of the magnetic
vector is unknown, but the contribution of these unknown quantities
to the $\Bl$ signal changes with viewing angle.  In other words, when
$\mu=\cos(\theta)=1.0$, the line-of-sight component $\Bl$ is
equal to the radial component $\Br$.  Nowhere else is this true.

\begin{figure}
\centerline{
\includegraphics[width=0.75\textwidth, clip, trim = 20mm 160mm 0mm 40mm]{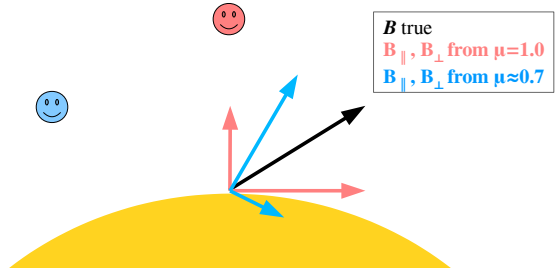}}
\caption{Geometry of projection effects when observing a photospheric magnetic
field vector (black) in the image plane, from two different perspectives.
Of note regarding the inferred polarity is that the $\Bl$ component (pointed
toward/away from the observer) changes sign between the $\mu=1.0$ and 
$\mu=0.7$ viewing angles.  }
\label{fig:projection}
\end{figure}

A common approach to alleviate some of this ``projection effect''
on the inferred total field strength is to assume that the field
vector is radial everywhere; then by dividing the observed $\Bl$
by $\mu$, an approximation to the radial field may be retrieved.
This is the ``$\mu$-correction''.  Its earliest uses first supported
the hypothesis of the overall radial nature of plage and polar fields,
and provided a reasonable estimate of polar fields for heliospheric and
coronal magnetic modeling \cite{Svalgaard_etal_1978,WangSheeley1992}.
However, it was evident from very early
studies using longitudinal magnetographs and supporting chromospheric
imaging that sunspots were composed of fields which were significantly
non-radial, {\it i.e.} inclined with respect to the local normal.  This geometry can lead to the notorious introduction of
apparent flux imbalance and ``false'' magnetic polarity inversion lines
(see Figure~\ref{fig:projection}) when the magnetic vector's inclination relative to the line-of-sight
surpasses $90^\circ$ while the inclination to the local vertical remains less than 
$90^\circ$ or {\it vice versa} \cite{ChapmanSheeley1968,PopeMosher1975,Giovanelli1980,Jones1985}.
Although this artifact can be cleverly used for some investigations
\cite{SainzDalda_MartinezPillet_2005} it generally poses a
hindrance to interpreting the inherent solar magnetic structure present.

The inaccuracies which arise from using $\Bl$ are generally assumed to
be negligible when the observing angle $\theta$ is less than $30^\circ$;
if the field is actually radial, the correction is only a $\approx13$\%
error, and introduced false neutral lines generally appear only in
the super-penumbral areas.   Yet this is a strong
assumption, and one known to be inaccurate for many solar magnetic
structures.  The estimates of polar radial field strength are especially
crucial for global coronal field modeling and solar wind estimations
\cite{pfss_mhd_comp,Riley_etal_2014}, but these measurements are
exceedingly difficult \cite{Tsuneta_etal_2008,Ito_etal_2010,Petrie_2015}. 

The gains afforded by using the full magnetic vector include the
ability to better estimate the $\Br$ component by way of a coordinate
transform of the azimuthally-disambiguated inverted Stokes vectors
\cite{garyhagyard90}.  However, although there are instruments now which
routinely provide full-disk vector magnetic field data (such as SOLIS,
\opencite{solis}; HMI \opencite{hmi_pipe}), the line-of-sight component
$\Bl$ remains the least noisy, easiest measurement of basic solar magnetic
field properties.

We present here a method of retrieving a different, and in the case of
sunspots, demonstrably better, estimate of the radial field boundary,
$\Brp$, the radial component of a potential field which is constructed
from $\Bl$ so as to match to the observed line-of-sight component.
This approach was originally described by \inlinecite{Sakurai1982}
and \inlinecite{Alissandrakis1981}, but is rarely used in the
literature.  We demonstrate here its implementation, including 
in spherical geometry for full-disk data.  The method is
described in section~\ref{sec:method}, the data used are described
in section~\ref{sec:data}, and both the planar and spherical results
for $\Brp$ are evaluated quantitatively for active-region and solar
polar areas in section~\ref{sec:arcomps} and \ref{sec:fluxcomps}.
In section~\ref{sec:rcomps} we reflect specifically on the different
approximations in the context of ``false polarity-inversion lines''
artifacts, and in section~\ref{sec:successNfailure} we investigate the reasons
behind both regions of success and areas of failure.

\section{Method}
\label{sec:method}

Approaches to computing the potential field which matches the observed
line-of-sight component are outlined here for two geometries, with details given in
Appendices~\ref{app:method_planar} and \ref{app:method_spherical}.  When one is
focused on a limited part of the Sun such that curvature effects are minimal, a
planar approach can be used to approximate the radial field (section
\ref{sec:method_planar}, appendix~\ref{app:method_planar}).  The planar
approach is fast, and reasonably robust for active-region sized patches
(section \ref{sec:arcomps}).  When the desired radial-field boundary is the
full disk, or covers an extended area of the disk, such as
the polar area, then the spherical extension of the method is the most
appropriate (section \ref{sec:method_spherical},
appendix~\ref{app:method_spherical}); however, depending on the image size of
the input, calculating the radial field in this way can be quite slow.

\subsection{Method: Planar Approximation}
\label{sec:method_planar}

The line-of-sight component is observed on an image-coordinate planar grid. We
avoid having to interpolate to a regular heliographic grid by performing the
analysis using a uniform grid in image coordinates, $(\xi, \eta)$.  Restricting
the volume of interest to $0<\xi<L_x$, $0<\eta<L_y$ and $z\ge0$, and neglecting
curvature across the field of view, the potential field can be written in terms
of a scalar potential $\BB^{\rm pot}=\grad \Phi$, with the scalar potential
given by 
\begin{eqnarray}
\Phi(\xi,\eta,z) &=& \sum_{m,n} A_{mn} 
e^{[2 \pi i m \xi/L_x + 2 \pi i n \eta/L_y - \kappa_{mn} z]} + A_0 z, 
\end{eqnarray} 
where $z$ is the vertical distance above the solar surface. 
The value of $\kappa_{mn}$ is determined by $\grad^2 \Phi=0$, namely 
\begin{eqnarray}
\kappa_{mn}^2 &=& (2 \pi)^2 \bigg [(c_{11}^2 + c_{12}^2) \bigg ({m \over L_x} \bigg )^2 
+ 2 (c_{11} c_{21} + c_{12} c_{22}) \bigg ({m \over L_x} \bigg ) \bigg ({n \over L_y} \bigg ) 
\nonumber \\
&& \qquad + (c_{21}^2 + c_{22}^2) \bigg ({n \over L_y} \bigg )^2 \bigg ],
\end{eqnarray} 
where $c_{ij}$ are the coordinate transformation coefficients given in
\inlinecite{garyhagyard90}, and choose $\kappa_{mn}>0$ so the field stays
finite at large heights.  The values of the coefficients $A_{mn}$ are
determined by requiring that the observed line-of-sight component of the field
matches the line-of-sight component of the potential field, which results in 
\begin{eqnarray}
{\tt FFT}(B^l) &=& L_x L_y A_{jk} \bigg [{2 \pi i j \over L_x} (c_{11} a_{13} + c_{12} a_{23}) 
+ {2 \pi i k \over L_y} (c_{21} a_{13} + c_{22} a_{23}) - \kappa_{jk} a_{33} \bigg ] 
\nonumber \\ 
&& \quad + L_x L_y a_{33} A_0 \delta_{0j} \delta_{0k}. 
\end{eqnarray} 
where $a_{ij}$ are the elements of the field components transformation matrix
given in \inlinecite{garyhagyard90}, and ${\tt FFT}(B^l)$ denotes taking the
Fourier Transform of $B^l$.  The value of $A_0$ is determined by the net
line-of-sight flux through the field of view.

Placing the tangent point at the center of the presented field of view
(corresponding to evaluating the coordinate transformation coefficients and the
field components transformation matrix at the longitude and latitude of the
center of the field of view) is the default, producing a boundary labeled
$\Bzppc$. The resulting radial component estimation is least accurate near the
edges, as expected, and as such we also present $\Brppa$, for which the tangent
point is placed at each pixel presented, the field calculated, and only that
pixel's resulting radial field (for which it acted as the tangent point) is
included.

\subsection{Method: Spherical Case}
\label{sec:method_spherical}

The potential field in a semi-infinite volume $r \ge R$ can be written in terms
of a scalar potential $\BB^{\rm pot}=-\grad \Psi$, given by
\begin{eqnarray} 
\Psi &=& R \sum_{n=1}^\infty \sum_{m=0}^n \bigg ({R \over r} \bigg )^{n+1} 
(g_n^m \cos m \phi + h_n^m \sin m \phi) P_n^m(\mu),
\end{eqnarray} 
where $\mu=\cos \theta$.  Defining the coordinate system such that the line of
sight direction corresponds to the polar axis of the expansion results in
particularly simple expressions for the coefficients $g_n^m$, $h_n^m$
\cite{Rudenko2001a}.  However, because observations are only available for the
near side of the Sun, it is necessary to make an assumption about the far side
of the Sun.  The resulting potential field at the surface $r=R$ is not
sensitive to this assumption except close to the limb, so for convenience, let
$B_l(R,\pi-\theta,\phi)=B_l(R,\theta,\phi)$, where the front side of the Sun is
assumed to lie in the range $0<\theta<\pi/2$.  

With these conventions, the coefficients are determined from 
\begin{eqnarray} 
g_n^m &=& \cases{{(2 n + 3) (n - m)! \over 2 \pi (n + m + 1)!} 
\int_0^{2\pi} d\phi \, \cos m \phi \int_0^1 d\mu \, P_{n+1}^m(\mu) 
B_l(R,\mu,\phi) & $n+m$ odd \cr
0 & $n+m$ even \cr}
\end{eqnarray} 
and 
\begin{eqnarray} 
h_n^m &=& \cases{{(2 n + 3) (n - m)! \over 2 \pi (n + m + 1)!} 
\int_0^{2\pi} d\phi \, \sin m \phi \int_0^1 d\mu \, P_{n+1}^m(\mu) 
B_l(R,\mu,\phi) & $n+m$ odd \cr
0 & $n+m$ even \cr}
\end{eqnarray} 
and the radial component of the field is given by 
\begin{eqnarray} 
B_r &=& - {\partial \Psi \over \partial r} 
= \sum_{n=1}^\infty \sum_{m=0}^n (n + 1) \bigg ({R \over r} \bigg )^{n+2} 
(g_n^m \cos m \phi + h_n^m \sin m \phi) P_n^m(\mu).
\end{eqnarray} 
When evaluated at $r=R$, this produces a full-disk radial field boundary
designated $\Brpsph$ from which extracted HARPs and polar sub-regions are
analyzed below.  Our implementation of this approach uses the Fortran 95
SHTOOLS library \cite{SHTOOLS_v3.4} for computing the associated Legendre
functions. The routines in this library are considered accurate up to degrees
of $n\approx2800$.  For the results presented here, a value of $n=2048$ was
used, corresponding to a spatial resolution of about 2\,Mm.

\section{Data}
\label{sec:data}

For this study we use solely the vector magnetic field observations from
the Solar Dynamics Observatory \cite{sdo} Helioseismic and Magnetic
Imager \cite{hmi,hmi_pipe}.  Two sets of data were constructed: a
full-disk test and a set of HMI Active Region Patches
(``HARPs''; \opencite{hmi_pipe}, \opencite{hmi_invert}, \opencite{hmi_sharps}) over 5 years.  For both,
in order to keep comparisons as informative as possible, we construct
line-of-sight component data from the vector data by transforming the
magnetic vector components into a $\Bl$ map: $\Bl = B \cos(\xi)$ where
$\xi$ is the inclination of the vector field in the observed plane of
the sky coordinate system, as returned from the inversion.

The full-disk data target is {\tt 2011.03.06\_15:48:00\_TAI}; this date
was chosen due to its extreme {\tt B0} angle such that the south
pole of the Sun is visible, the variety of active regions visible at
low $\mu=\cos(\theta)$ observing angle (away from disk center), and the
presence of a northern extension of remnant active-region plage.  The {\tt
hmi.ME\_720s\_fd10} full-disk series was used; data are available through the JSOC
lookdata tool\footnote{jsoc.stanford.edu/lookdata.html}.  Because the
weaker fields do not generally have their inherent $180^\circ$ ambiguity
resolved in that series and we will be evaluating the
$\Brp$ method in poleward areas of weaker field, two customizing steps
were taken.  First, a custom noise mask was generated ($\textsc{ambthrsh}=0$,
rather than the default value of 50).  Second a custom disambiguation
was performed using the cooling parameters: $\textsc{ambtfctr}=0.998$, 
$\textsc{ambneq}=200$, $\textsc{ambngrow}=2$, 
$\textsc{ambntx}=\textsc{ambnty}=48$ \cite{hmi_pipe,hmi_ambig};
compared to the default HMI pipeline implementation, these parameters
provide smaller tiles over which the potential field is computed
to estimate $d{\BB}/d{\rm z}$, a smaller ``buffer'' of noisy pixels
around well-determined pixels, and slower cooling for the simulated annealing
optimization.  Disambiguation results were generated
for 10 random number seeds.  Pixels used for the comparisons shown
herein are only those for which both the resulting equivalent of the
$\textsc{conf\_disambig}$ segment is $\geq$60 and the results from all 10
random number seeds agreed, as well.   This requirement translates to
75.3\% of the pixels with $\textsc{conf\_disambig} \ge 60$ and 88.8\% of
the pixels with $\textsc{conf\_disambig}=90$ being included.  In the case
of the present data, there is a 0.2\% chance that the disambiguation
solution used results by chance, even in the weak areas (including the poles).

\begin{figure}
\centerline{
\includegraphics[width=0.75\textwidth, clip, trim = 0mm 0mm 0mm 0mm]{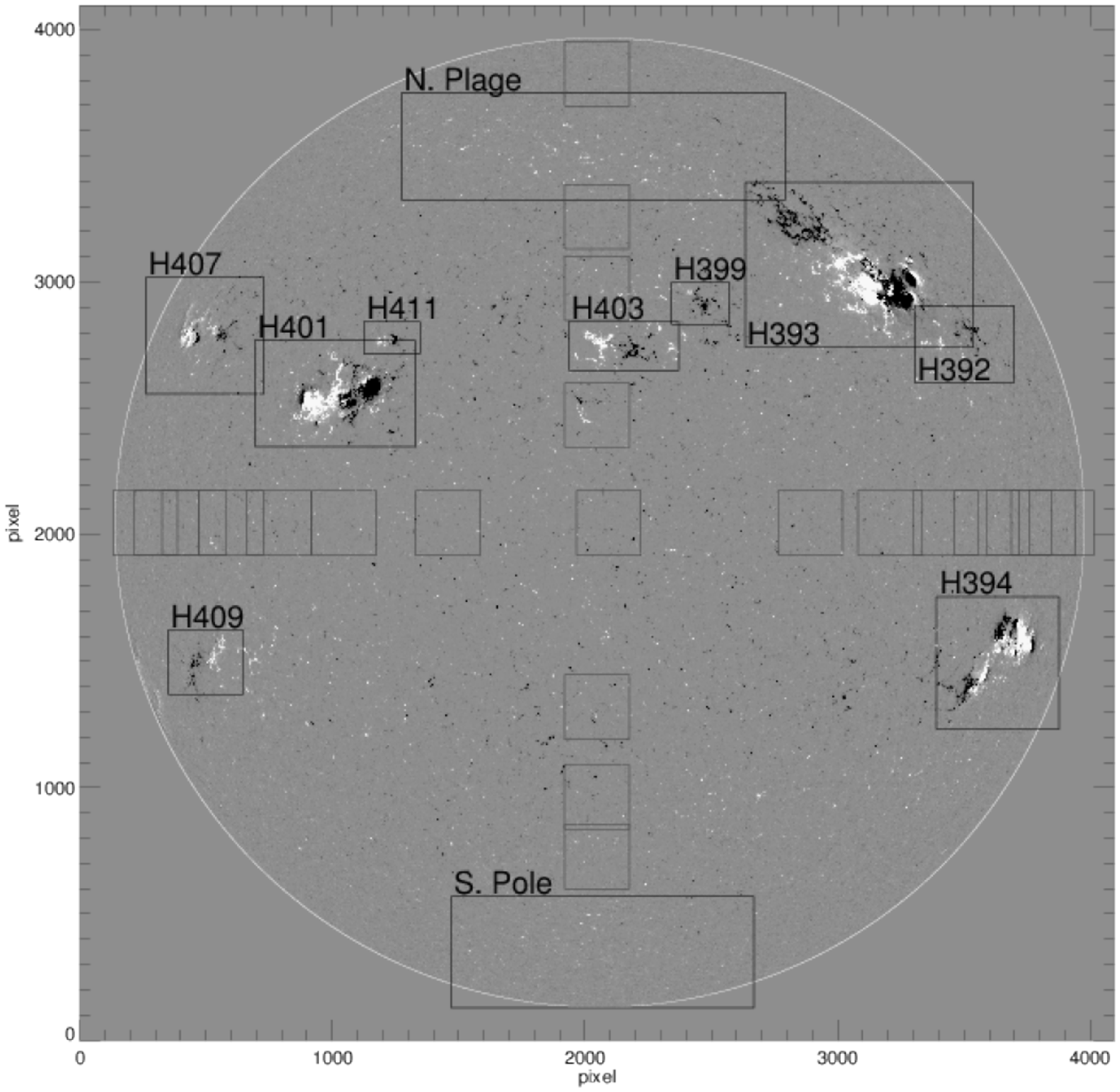}}
\caption{Full-disk image of the line-of-sight component of the solar photospheric
magnetic field on 2011.03.06 at 15:48:00\_TAI, scaled to $\pm 200$G.  The solar 
limb is indicated as are the HMI Active Region Patches labeled with their
``HARP number'', and the additional two polar areas used for this analysis.  Additional
small patches of quiet-Sun, distributed in $\mu=\cos{\theta}$ are indicated as grey
boxes, occasionally overlapping.  Solar
north is up, west to the right, positive/negative directed field is shown as white/black
respectively.}
\label{fig:harps}
\end{figure}

From this full disk magnetogram the nine identified HMI
Active Region Patch (``HARP'') areas are extracted using the keywords $\textsc{harpnum}$,
$\textsc{crpix1}$, $\textsc{crpix2}$, $\textsc{crsize1}$, $\textsc{crsize2}$ from the {\tt hmi.Mharp\_720s} series.
Two polar regions were also extracted: a ``Northern plage
area'', which is an extended remnant field and a ``South Pole Region''
that encompasses the entire visible polar area.  A context image is shown
in Figure~\ref{fig:harps}, and summary information about each sub-area
is given in Table~\ref{table:harps}, including the WCS coordinates 
for the two non-HARP regions for reproducibility.  
Also in Table~\ref{table:harps} are summary data for an additional 22 sub-areas,
each $256^2$ pixels in size centered along the midpoints in $x,~y$ on the image
at a variety of $\mu=\cos{\theta}$ positions.

\begin{table}
\caption{Extracted Area Descriptions} 
\begin{tabular}{cccccl}
I.D. & HARP  & NOAA  & Locale & $\mu = \cos(\theta)$ & Description \\ 
     & No. & AR No. & &  & \\ \hline
{\tt H392} & 392 & 11163 & N22 W54 & 0.54 & small plage \\
{\tt H393} & 393 &11164 & N32 W39& 0.65  & large complex active region\\
{\tt H394} & 394 & 11165 & S17 W59 & 0.49 &  small active region \\ 
{\tt H399} & 399 & N/A  & N27 W13 & 0.87 &  small plage \\
{\tt H401} & 401 & 11166 & N15 E34 & 0.79 & simple active region \\
{\tt H403} & 403 & 11167 & N21 W03 & 0.93 &  bipolar plage \\
{\tt H407} & 407 & 11169 & N23 E62 & 0.43 & small active region \\
{\tt H409} & 409 & N/A & S17 E58 & 0.51 & small plage \\
{\tt H411} & 411 & N/A & N22 E27 & 0.92 & small spot \\ 
{\tt N.Plage} & N/A & N/A & N51 E01 & 0.63 & north remnant plage\\
 & & & & & $\textsc{crpix1}$=1304 $\textsc{crpix2}$=344 \\
 & & & & & $\textsc{crsize1}$=1516 $\textsc{crsize2}$=428 \\
{\tt S.Pole} & N/A & N/A & S60 E01 & 0.51 & south polar area \\
 & & & & & $\textsc{crpix1}$=1199 $\textsc{crpix2}$=3527 \\
 & & & & & $\textsc{crsize1}$=1427 $\textsc{crsize2}$=440 \\ 
{\tt QS\_E\_350} & N/A & N/A & N00 E70 & 0.35 & $\textsc{crpix1}$=3709 $\textsc{crpix2}$=1921 \\
{\tt QS\_E\_450} & N/A & N/A & N00 E63 & 0.45 & $\textsc{crpix1}$=3625 $\textsc{crpix2}$=1921 \\
{\tt QS\_E\_550} & N/A & N/A & N00 E57 & 0.55 & $\textsc{crpix1}$=3514 $\textsc{crpix2}$=1921 \\
{\tt QS\_E\_650} & N/A & N/A & N00 E49 & 0.65 & $\textsc{crpix1}$=3370 $\textsc{crpix2}$=1921 \\
{\tt QS\_E\_750} & N/A & N/A & N00 E41 & 0.35 & $\textsc{crpix1}$=3181 $\textsc{crpix2}$=1921 \\
{\tt QS\_E\_850} & N/A & N/A & N00 E32 & 0.85 & $\textsc{crpix1}$=2923 $\textsc{crpix2}$=1921 \\
{\tt QS\_E\_950} & N/A & N/A & N00 E18 & 0.95 & $\textsc{crpix1}$=2511 $\textsc{crpix2}$=1921 \\
{\tt QS\_W\_1000} & N/A & N/A & N00 W01 & 1.00 & $\textsc{crpix1}$=1874 $\textsc{crpix2}$=1921 \\
{\tt QS\_W\_900} & N/A & N/A & N00 W26 & 0.90 & $\textsc{crpix1}$=1076 $\textsc{crpix2}$=1921 \\
{\tt QS\_W\_800} & N/A & N/A & N00 W37 & 0.80 & $\textsc{crpix1}$=761 $\textsc{crpix2}$=1921 \\
{\tt QS\_W\_700} & N/A & N/A & N00 W46 & 0.70 & $\textsc{crpix1}$=542 $\textsc{crpix2}$=1921 \\
{\tt QS\_W\_600} & N/A & N/A & N00 W53 & 0.60 & $\textsc{crpix1}$=377 $\textsc{crpix2}$=1921 \\
{\tt QS\_W\_500} & N/A & N/A & N00 W60 & 0.50 & $\textsc{crpix1}$=251 $\textsc{crpix2}$=1921 \\
{\tt QS\_W\_400} & N/A & N/A & N00 W66 & 0.40 & $\textsc{crpix1}$=154 $\textsc{crpix2}$=1921 \\
{\tt QS\_W\_300} & N/A & N/A & N00 W72 & 0.30 & $\textsc{crpix1}$=82 $\textsc{crpix2}$=1921 \\
{\tt QS\_N\_375} & N/A & N/A & N68 E01 & 0.375 & $\textsc{crpix1}$=1921 $\textsc{crpix2}$=143 \\
{\tt QS\_N\_775} & N/A & N/A & N39 E00 & 0.775 & $\textsc{crpix1}$=1921 $\textsc{crpix2}$=708 \\
{\tt QS\_N\_875} & N/A & N/A & N29 E00 & 0.875 & $\textsc{crpix1}$=1921 $\textsc{crpix2}$=992 \\
{\tt QS\_N\_975} & N/A & N/A & N13 E00 & 0.975 & $\textsc{crpix1}$=1921 $\textsc{crpix2}$=1495 \\
{\tt QS\_S\_925} & N/A & N/A & S22 E00 & 0.925 & $\textsc{crpix1}$=1921 $\textsc{crpix2}$=2650 \\
{\tt QS\_S\_825} & N/A & N/A & S34 E00 & 0.825 & $\textsc{crpix1}$=1921 $\textsc{crpix2}$=3005 \\
{\tt QS\_S\_725} & N/A & N/A & S44 E00 & 0.725 & $\textsc{crpix1}$=1921 $\textsc{crpix2}$=3242 \\ \hline
\label{table:harps} 
\end{tabular}
\end{table}

Throughout this study, we differentiate between when planar approximations
are invoked and when curvature is accounted for by referring to 
``$\Bz$'' and ``$\Br$'', respectively.
The potential-field approximation is performed three ways, as described
in section~\ref{sec:method}, above: using a planar approximation from the
center-point coordinates $\Bzppc$ of each HARP or extracted sub-region,
using a planar approximation with each point of the extracted region used
as the center point $\Brppa$, and using the spherical full-disk approach
$\Brpsph$.  In addition,
we calculate two common $\mu$-correction approximations
for each sub-region: the center point value of $\mu=\cos ( \theta )$
used as a tangent point to obtain $\Bzm = \Bl / \mu$, and secondly
each pixel's $\mu$ value is calculated and applied independently, for
$\Brms = \Bl/\mu(s)$ where $(s)$ is the spatial location of the pixel.

The second set of data consists of a subset of all HARPs selected over
5.5 years, selected so as to generally not repeat sampling any particular
HARP: on days ending with '5' (5th, 15th, and 25th) of all
months 2010.05 -- 2015.06, the first `good' (quality flag is 0) HARP set at {\tt :48} past each
hour on/after {\tt 15:48} was used.  HARPs which were defined but for which
there were no active pixels are skipped.  The result is 1,819 extracted
HARPs without regard for size, complexity, or location on the disk.
Effectively the {\tt hmi.Bharp\_720s} series was used, including the
standard pipeline disambiguation.  NWRA's database initially began
construction prior to the pipeline disambiguation being performed for
earlier parts of the mission, thus for some of the data base, the  {\tt
hmi.ME\_720s\_fd10} data were used, the HARP regions extracted and the
disambiguation performed in-house, matching the implementation performed
in the HMI pipeline.  All analysis is performed up to
$80^\circ$ from disk center, and only points with significant signal/noise 
in the relevant components ($S/N>3$ relative to the returned uncertainties
from the inversion, and propagated accordingly) are included
in the analyses.  For this larger dataset, all calculations
are done with a planar approximation. The ``answer'' is $\Bz$,
and the boundary estimates calculated are: $\Bzm = \Bl / \mu$, $\Brms
= \Bl/\mu(s)$ (which imparts a spherical accounting due to the variation
of $\mu$ over the field of view), and $\Bzppc$.  $\Brppa$ and $\Brpsph$ are computationally 
possible but extremely slow, and are not employed for this second dataset.

\section{Results}

For the results presented here, the ``golden standard'' is taken to
be the radial or normal field as computed from the vector data, 
and to this quantity we compare results of different approximations of the boundary.
We do caution that $\Br$ data do include the observed $\Bt$
component, which is inherently noisier than the $\Bl$ component.
Additionally we stress that the comparisons are performed against
a particular instrument's retrieval of the photospheric magnetic field
vector, which may not reflect the true Sun as per influences in polarimetric
sensitivity, spectral finesse, spatial resolution, {\it etc.}

\subsection{Field Strength Comparisons}
\label{sec:arcomps}

\begin{figure}
\centerline{
\includegraphics[width=0.45\textwidth, height=0.35\textwidth, clip, trim = 15mm 0mm 5mm 10mm]{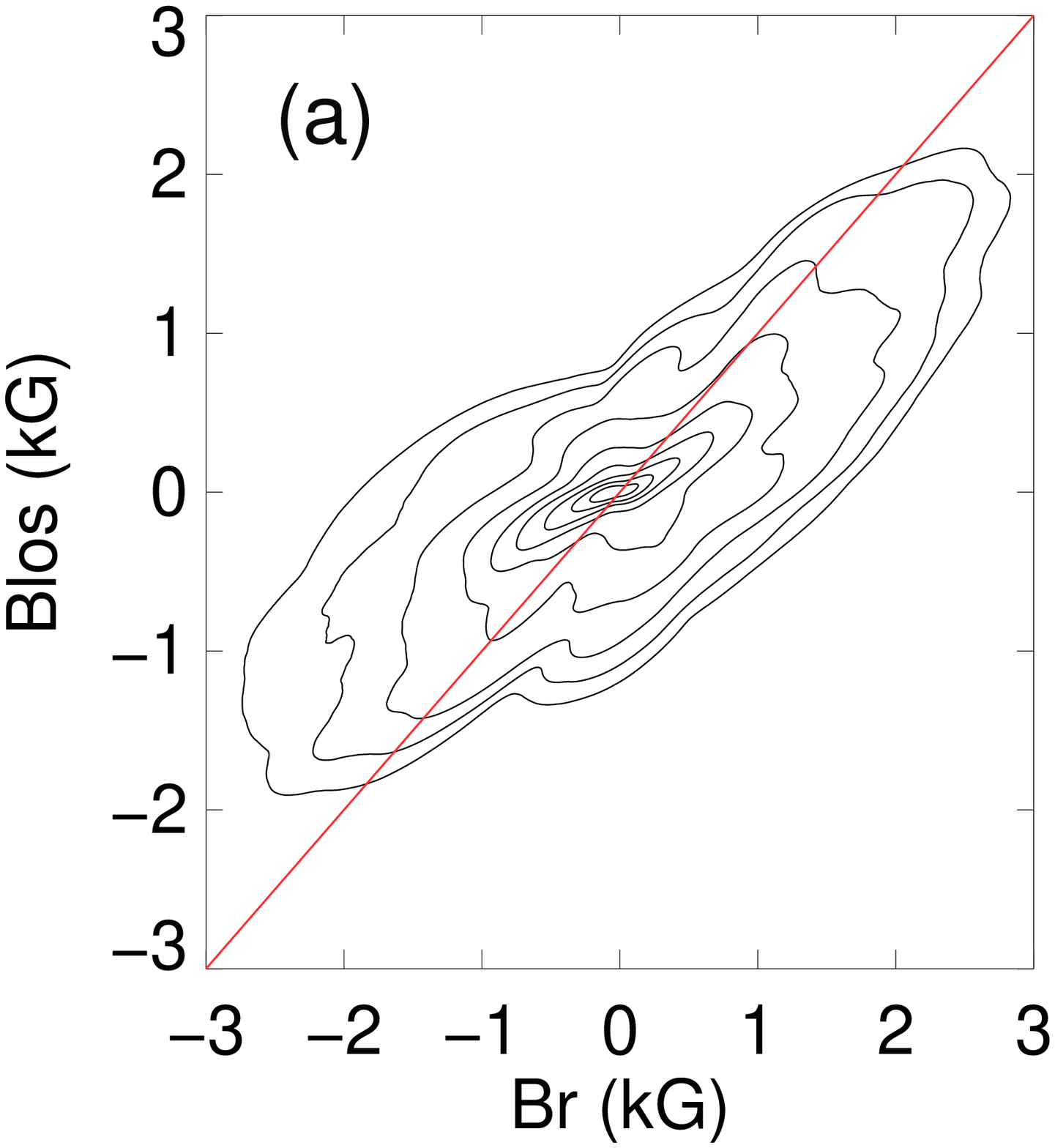}
\includegraphics[width=0.45\textwidth, height=0.35\textwidth, clip, trim = 15mm 0mm 5mm 10mm]{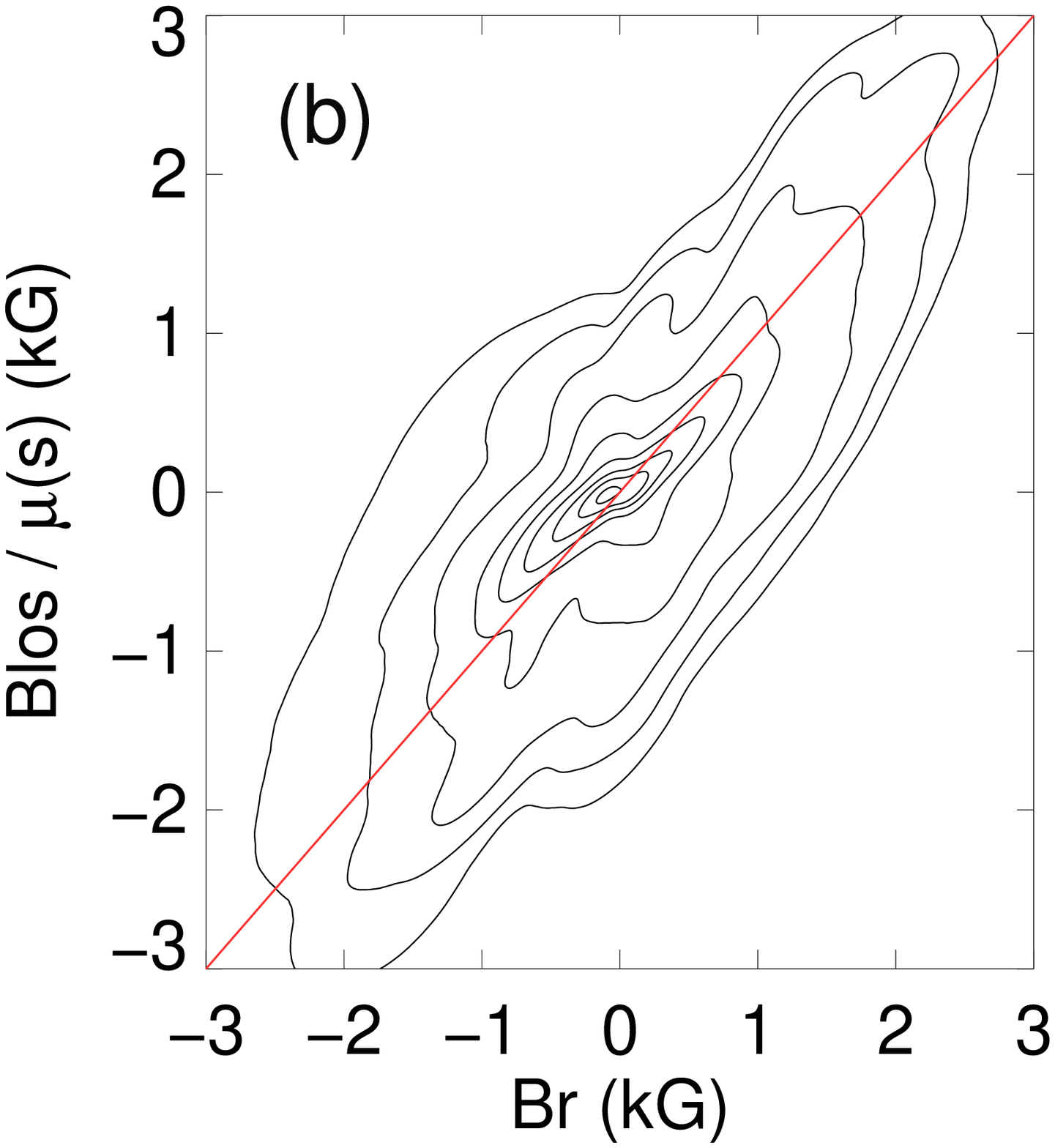}}
\centerline{
\includegraphics[width=0.45\textwidth, height=0.35\textwidth, clip, trim = 15mm 0mm 5mm 10mm]{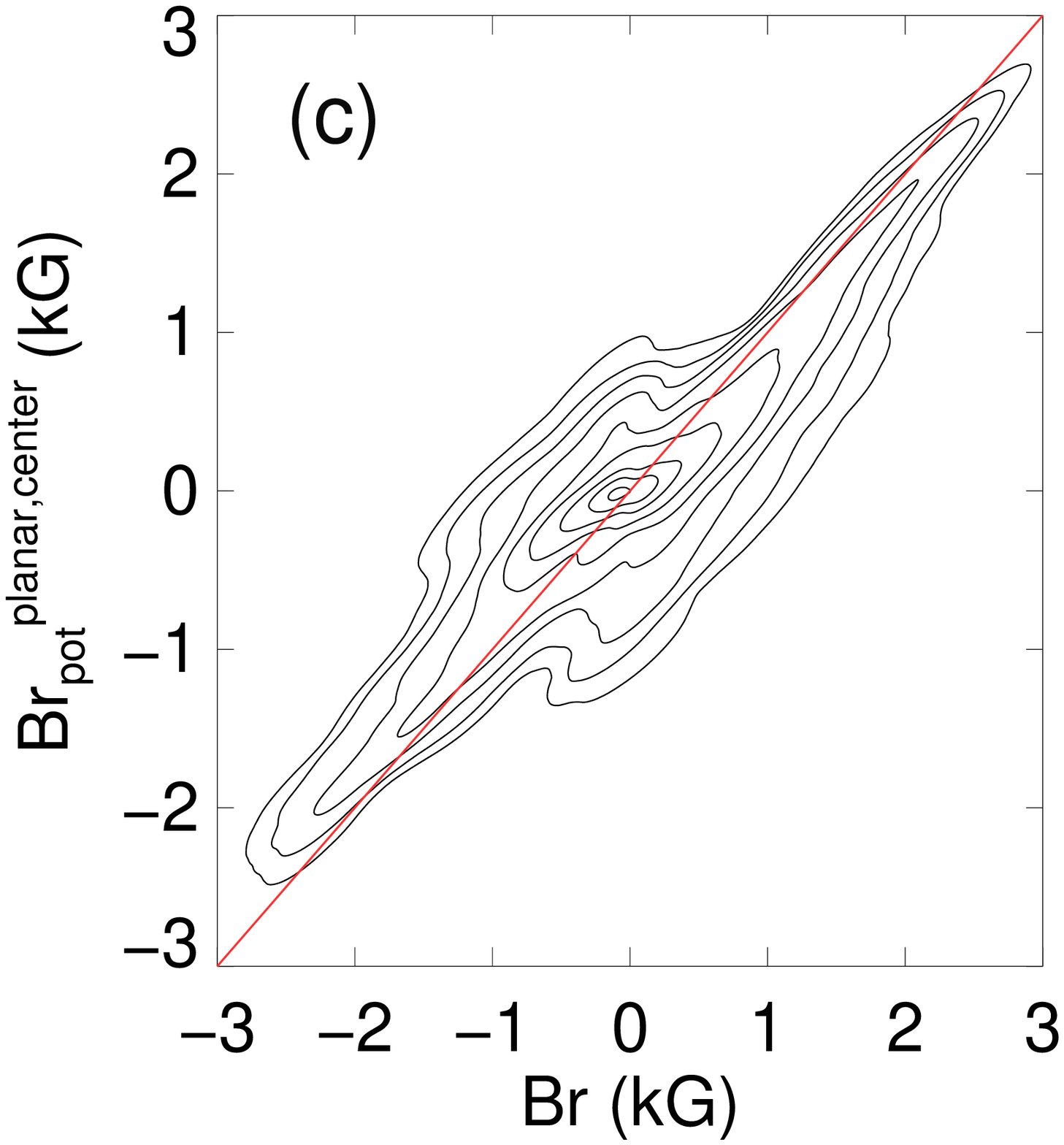}
\includegraphics[width=0.45\textwidth, height=0.35\textwidth, clip, trim = 15mm 0mm 5mm 10mm]{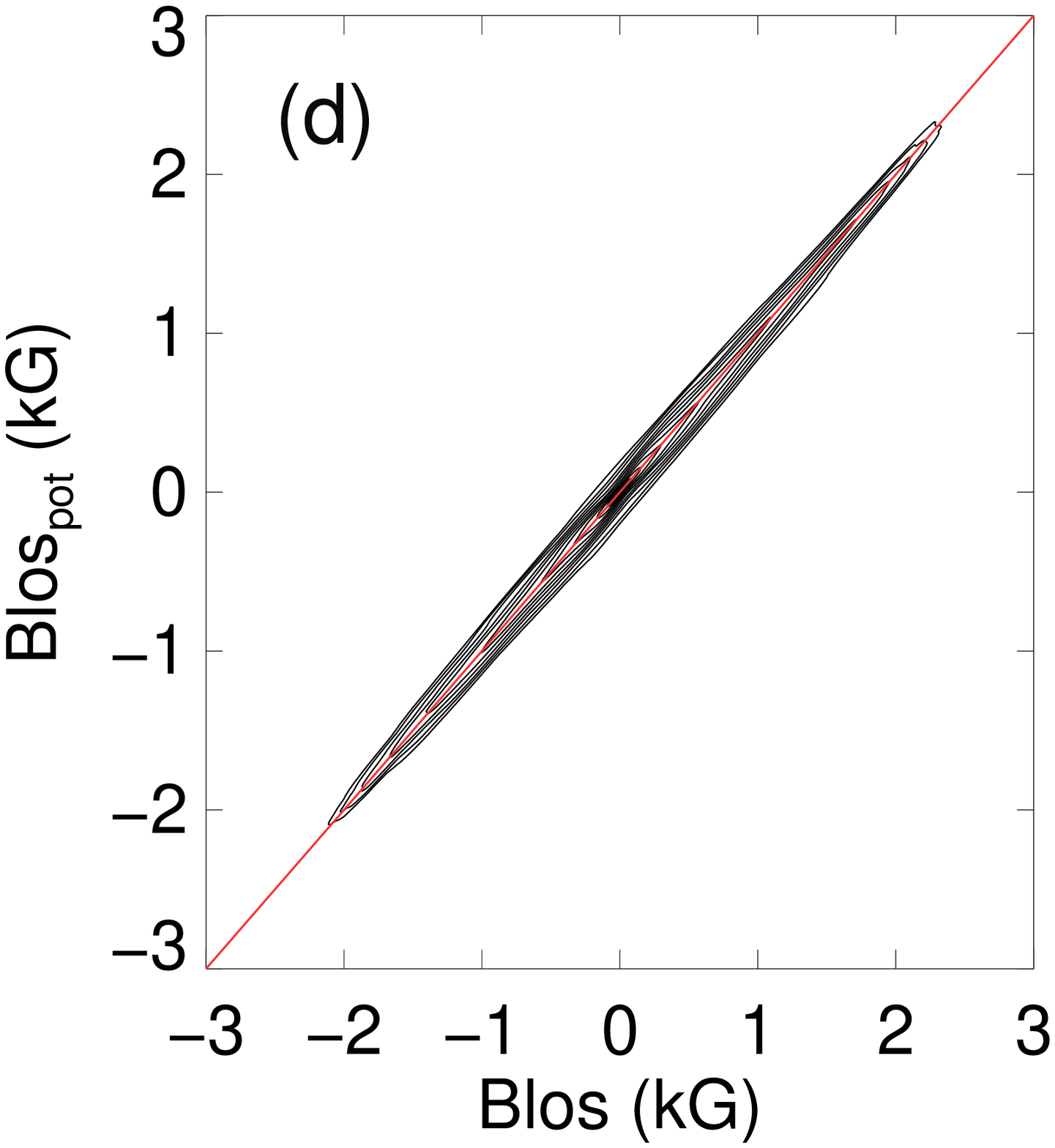}}
\caption{{\bf (a)--(c)} 
Non-parametric density estimates comparing the distribution of the 
radial field from the vector field data $\Br$ (x-axis) to the inferred radial field strength estimates 
using representatives of the different models discussed in the text (y-axis).
Shown here are distributions from NOAA\,AR\,11164 ({\tt H393}).  
Contour levels are equal in log (probability) ranging from [$10^{-3} - 10^{2}$], 
with $x=y$ line included for reference.
The comparisons are for $\Br$ against (a) $\Bl$ 
(b) $\Bl / \mu (s)$, 
(c) $\Bz$ potential, planar, center-point pivot
Panel (d) shows, in summary, that the $\mu$-corrected estimates 
generally show less bias compared to the $\Bl$ fields, but still
have a large random error. The potential field corrected fields show larger differences
for weak field strengths, but less random error.
In {\bf (d)} the scatter plot is between the $\Bl$ calculated from the inversion, and 
the recovered $\Bl$ 
calculated from the spherical potential vector (Sect.~\ref{app:method_spherical}).
While the boundary is thus fairly well recovered, the disagreement beyond machine-precision 
differences is due to the spherical calculation
not being performed with high enough degree to remove all small-scale ``ringing''.
}
\label{fig:scat393}
\end{figure}

\begin{figure}
\centerline{
\includegraphics[width=0.45\textwidth, height=0.35\textwidth, clip, trim = 15mm 0mm 5mm 10mm]{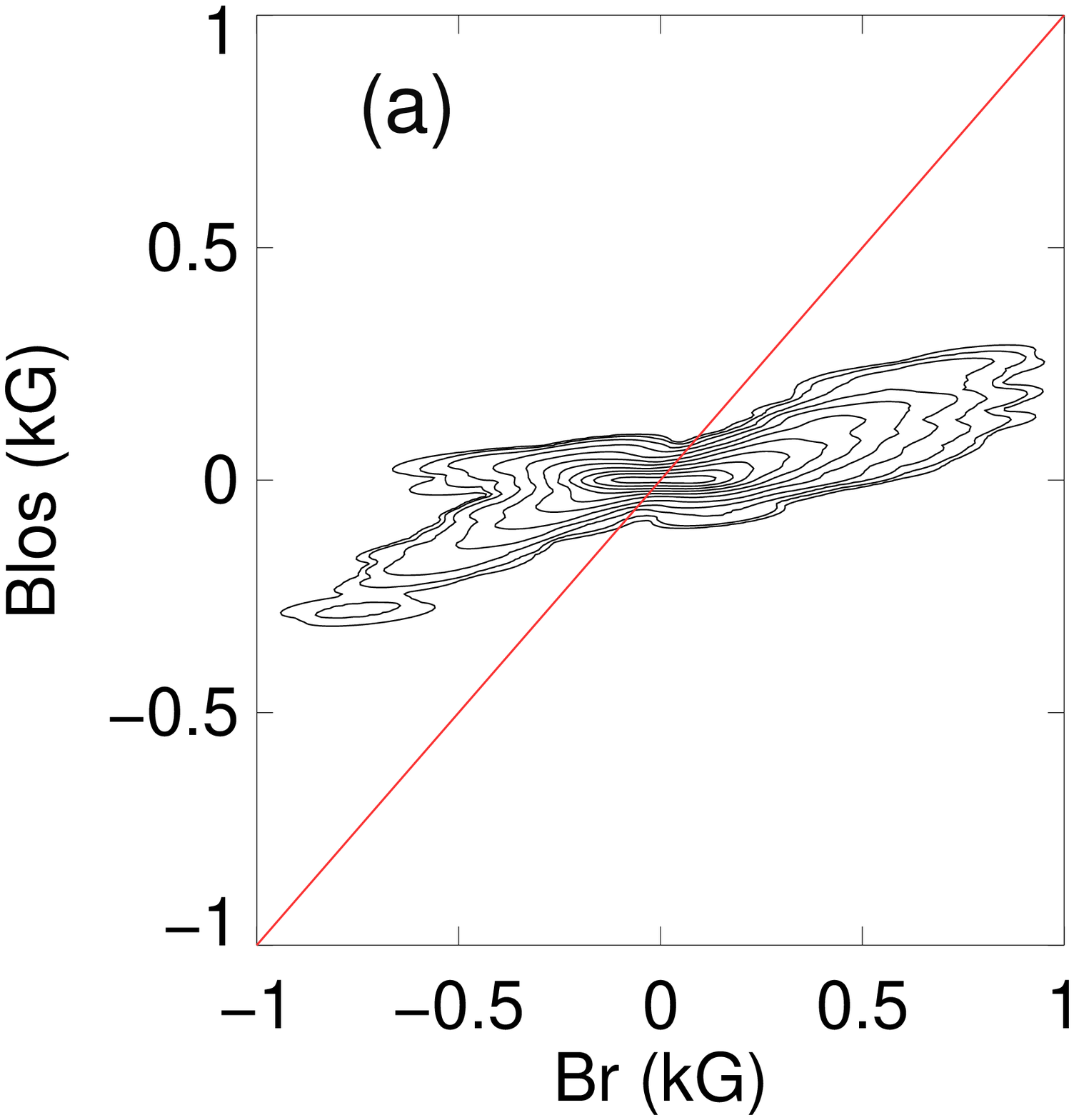}
\includegraphics[width=0.45\textwidth, height=0.35\textwidth, clip, trim = 15mm 0mm 5mm 10mm]{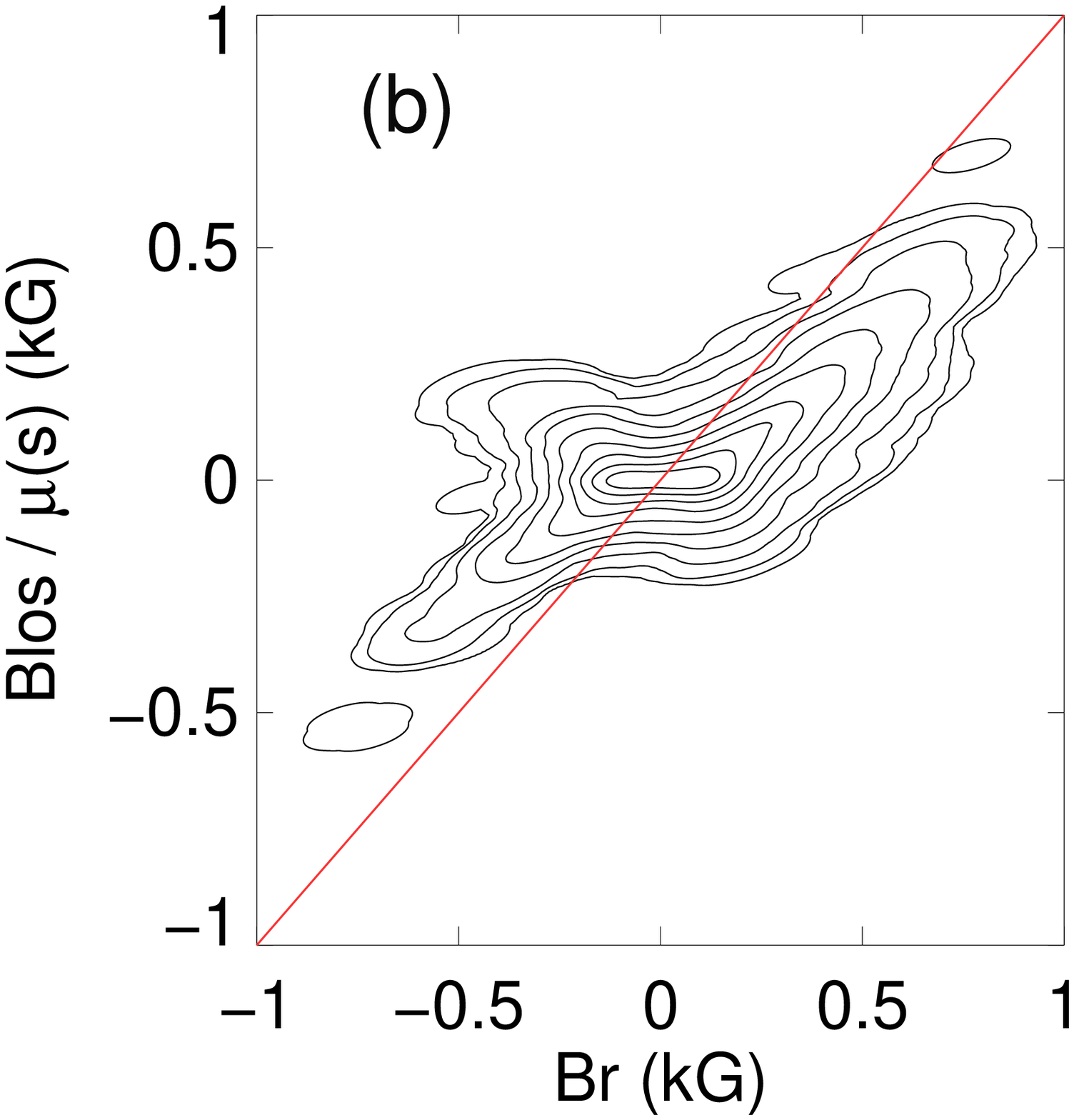}}
\centerline{
\includegraphics[width=0.45\textwidth, height=0.35\textwidth, clip, trim = 15mm 0mm 5mm 10mm]{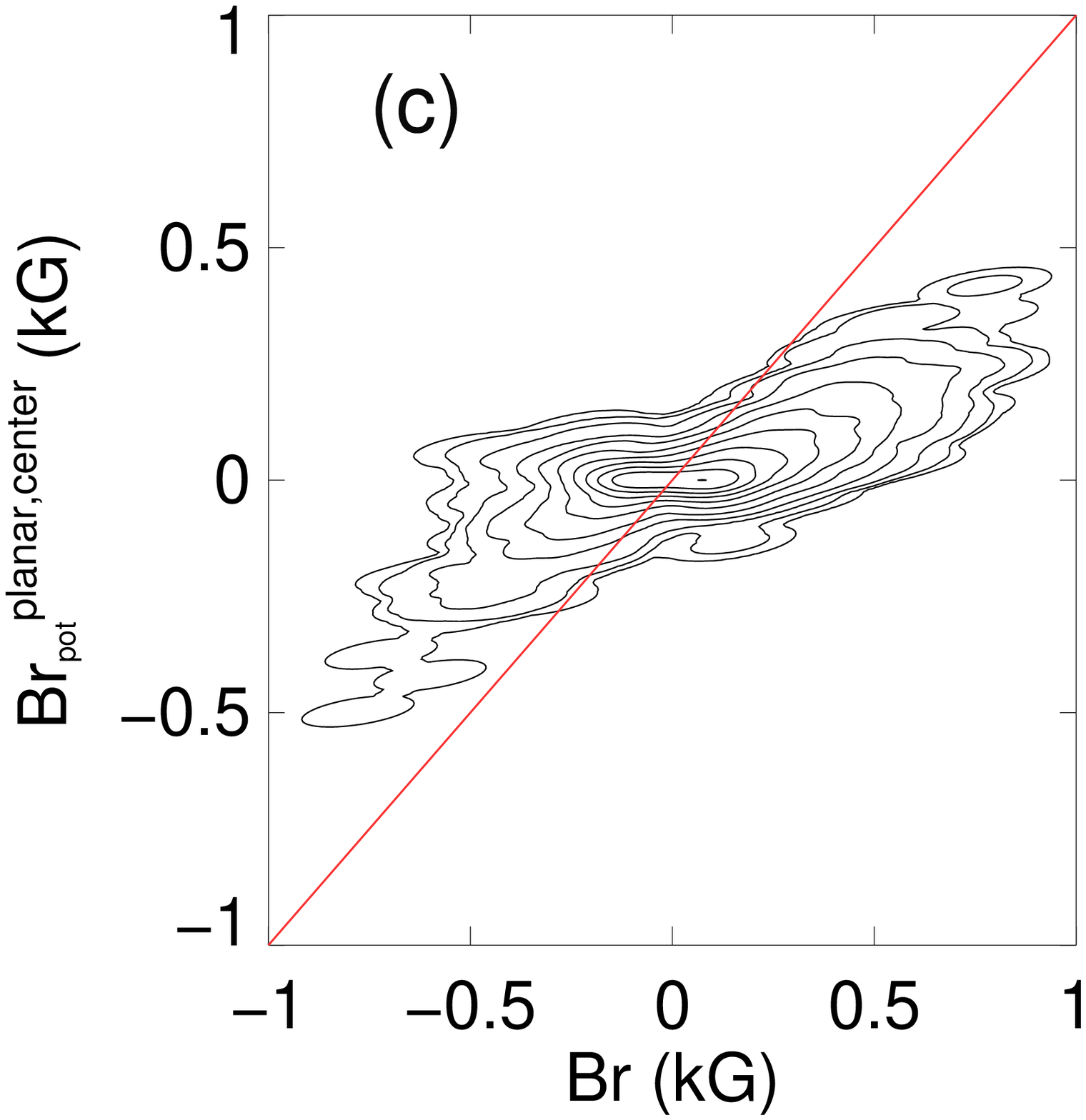}
\includegraphics[width=0.45\textwidth, height=0.35\textwidth, clip, trim = 15mm 0mm 5mm 10mm]{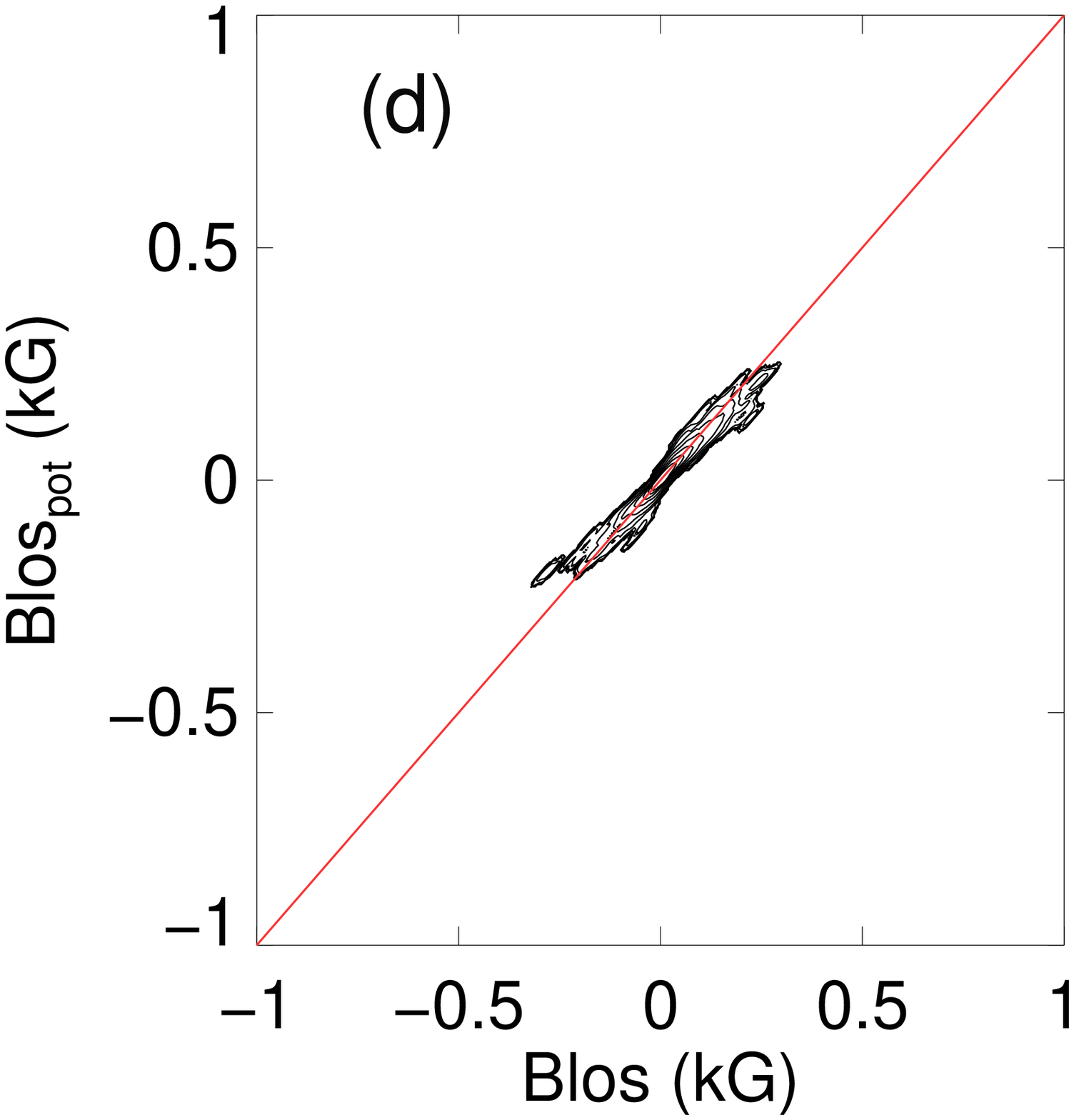}}
\caption{Same as Figure~\ref{fig:scat393}, but for the south pole area shown in 
Figure~\ref{fig:harps}.
These panels show, in summary, that the $\mu$-corrected estimates
for the south polar area show distinct spurs with incorrect polarity while the
potential field estimates have large biases.
}
\label{fig:scatspole}
\end{figure}

To demonstrate the general resulting trends for each of the 
radial field approximations, we first present density-histograms of
the inferred radial field strengths for two representative sub-regions,
NOAA\,AR\,11164 (HARP\,\#393, Figure~\ref{fig:scat393}) and
the south pole region (Figure~\ref{fig:scatspole}).
Throughout, we do not indicate the errors for clarity;
a 10\% uncertainty in field strength is a fair approximation overall,
and a detailed analysis beyond that level
is not informative here.

For both HARP {\tt H393} (NOAA\,AR\,11164) and the south pole area, the
initial comparison of $\Bl$ to $\Br$ (panel (a) in 
Figures~\ref{fig:scat393},~\ref{fig:scatspole}) shows the expected signature
of underestimated field strengths overall.  Note that for both regions,
but especially for the the south pole, there is a strong underestimation 
of the radial field strength across magnitudes.

The $\Brms$ correction (panel (b) in
Figures~\ref{fig:scat393},~\ref{fig:scatspole}; the $\Bzm$ plot is
almost identical and not shown here) shows improvement by eye for 
both regions, with distributions 
systematically deviating less from the $x=y$ line.  However, in the case
of the {\tt H393} corrections, the stronger-field strengths are often
over-corrected, and the opposite-polarity erroneous pixels are exacerbated
in their error.  This trend is also true for the south pole area: both
an improvement (especially for stronger-field points) and the
appearance of a distinct erroneous opposite-polarity spur in the $\Brms$ results.

In the next panel (panel (c) in Figures~\ref{fig:scat393},~\ref{fig:scatspole}), a 
representative potential-field option, $\Bzppc$,
is shown with regards to $\Br$ (again, $\Brpsph$ and $\Brppa$ plots look
essentially identical); the weaker field strengths appear to be less-well
corrected than the $\Brms$ approximation, although the strong-field
approximations for the sunspots in {\tt H393} are significantly better
than the $\Brms$ correction.  There appears to be a small number of
points for which the opposite polarity is retained but overall the
stronger field points (generally $\geq 1000$\,G) lie close to the $x=y$
line.  There are still significant deviations from the $\Br$ field; this is expected
at some level since a potential-field model is being imposed, and it cannot be expected 
that the solar magnetic fields are in fact potential.  Additionally, while 
a non-linear force-free field model may better represent the true field 
\cite{Livshits_etal_2015}, there is insufficient information in the $\Bl$
boundary with which to construct such a model.
For the south pole region, the strong-field areas are less well
corrected than was seen in the $\Brms$ plot (Figure~\ref{fig:scatspole}, panels c, b 
respectively), but the distinct
incorrect-polarity spur visible for $\Brms$ is less pronounced 
in the potential-field-based estimate.

For completeness, and as a check of the algorithm, the $\Bl$ directly
attained from the inversion as $\Bl = \vert \BB \vert \cos(\xi)$
where $\xi$ is the inclination of the field vector to the line
of sight (and which constitutes the input to the potential field
calculation), is compared with the $\Bl^{\rm pot}$ derived from
the vector field components from the derived potential field (panel
(d) in Figures~\ref{fig:scat393},~\ref{fig:scatspole}).  These $\Bl$
boundaries match well, indicating that there is little if any systematic
bias presented by the potential field calculation when recovering the
input boundary.  The recovery is not, however, within machine precision
due to the lower than optimal degree to which the spherical potential
field is computed; to reproduce the boundary to machine precision is
computationally untenable with this algorithm.

The two examples shown in Figures~\ref{fig:scat393},~\ref{fig:scatspole}
represent the two extremes of solar magnetic features to which these 
approximations would be applied: the south polar region (expected to sample
small, primarily radially-directed concentrations of field) and a large
active region with both plage and sunspots.  The distributions of the other 
sub-regions appear as hybrids when examined in the same manner, having
sometimes stronger fields for which the 
$\mu$-corrections approaches perform the best ({\it e.g.}, the northern plage 
area), or having sunspot areas for which there is an incorrect-polarity
spur present in the distributions that is exacerbated by some amount in 
the $\mu$-corrections and mitigated by some amount with the 
potential-field calculations.

\begin{figure}
\centerline{
\includegraphics[width=1.0\textwidth, clip, trim = 4mm 0mm 0mm 15mm]{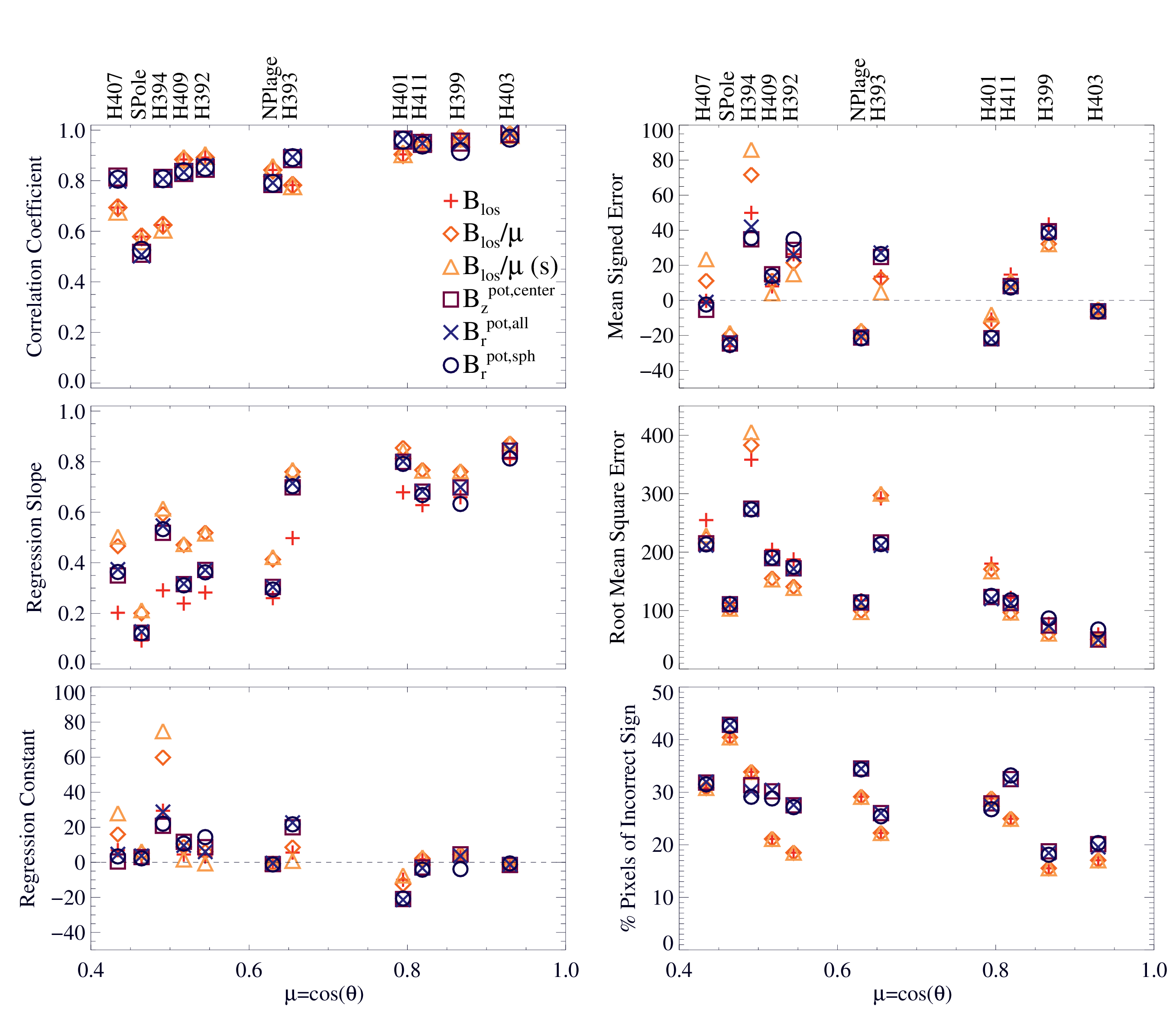}}
\caption{Metrics to evaluate the performance of radial-field approximations when compared
to $\Br$, for the cut-out areas highlighted in Figure~\ref{fig:harps}.  Top to Bottom, Left: 
the linear correlation coefficient, the slope of the linear regression line, 
the constant (offset) for that fit, and Right: the mean error, the root mean square error, 
the percent of pixels which are of the incorrect sign.
In shades of orange, $+$: $\Bl$, $\Diamond$: $\Bzm$, $\triangle$: $\Brms$; in shades
of purple, $\Box$: $\Bzppc$, $X$: $\Brppa$, $\bigcirc$: $\Brpsph$.
These metrics demonstrate some trends (less spread between methods with increasing $\mu$)
but also a mix of results between regions, $\mu$ and methods, indicating no single best approach.
}
\label{fig:stats_all}
\end{figure}

\begin{figure}
\centerline{
\includegraphics[width=1.00\textwidth, clip, trim = 4mm 0mm 0mm 15mm]{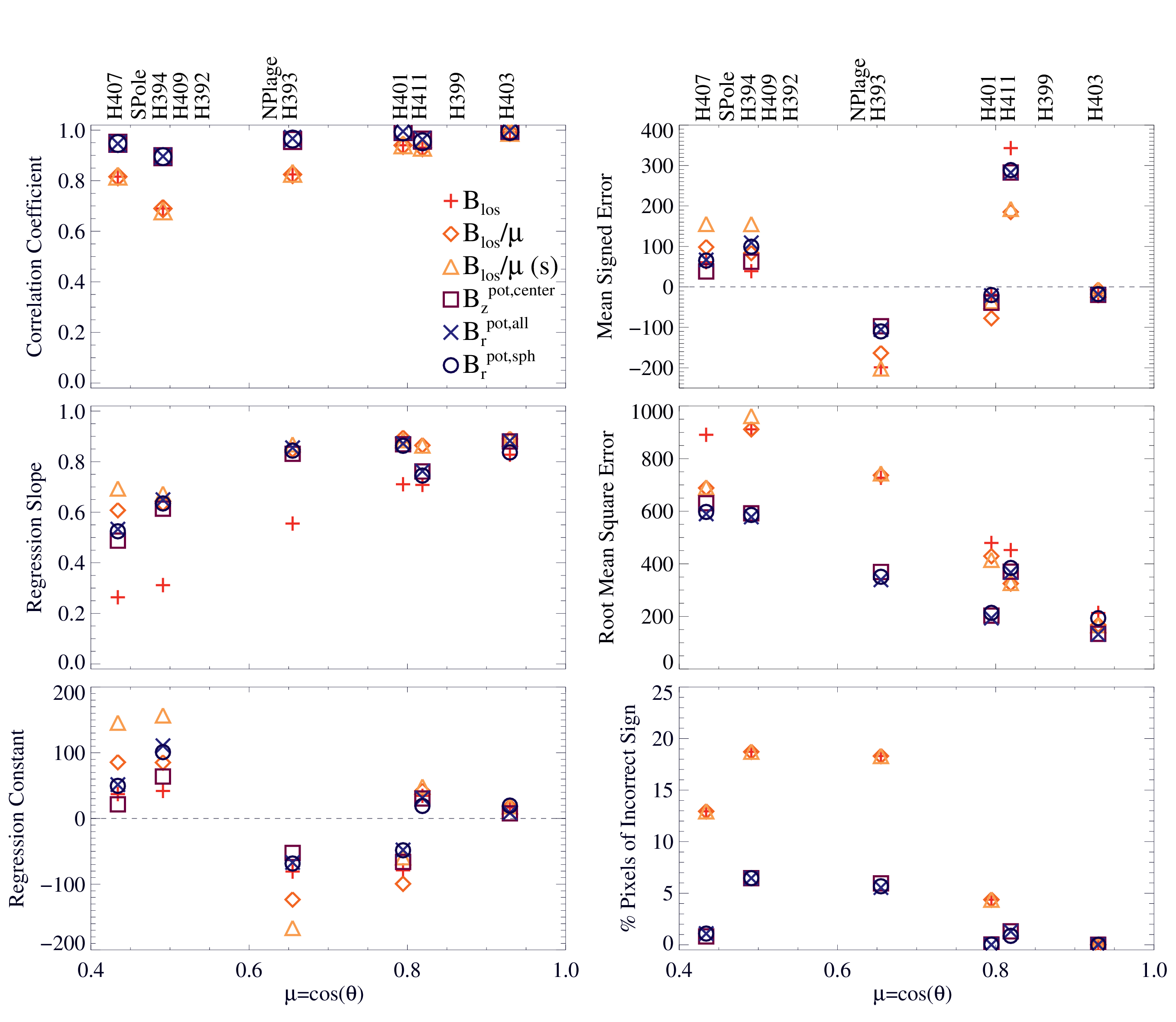}}
\caption{Same as Figure~\ref{fig:stats_all} but only for points with total field strengths
$|\BB| > 1000$G (note the different scales from prior figures for some metrics).  As some of the sub-regions do not meet the additional restriction of 
having a minimum of 100 such points, they are not included in this plot.  Note that 
compared with Fig.~\ref{fig:stats_all}, there are in some cases stronger distinctions
between the method categories, and much weaker relationships with 
$\mu=\cos(\theta)$ for some metrics.}
\label{fig:stats_strong}
\end{figure}

\begin{figure}
\centerline{
\includegraphics[width=1.00\textwidth, clip, trim = 4mm 0mm 0mm 15mm]{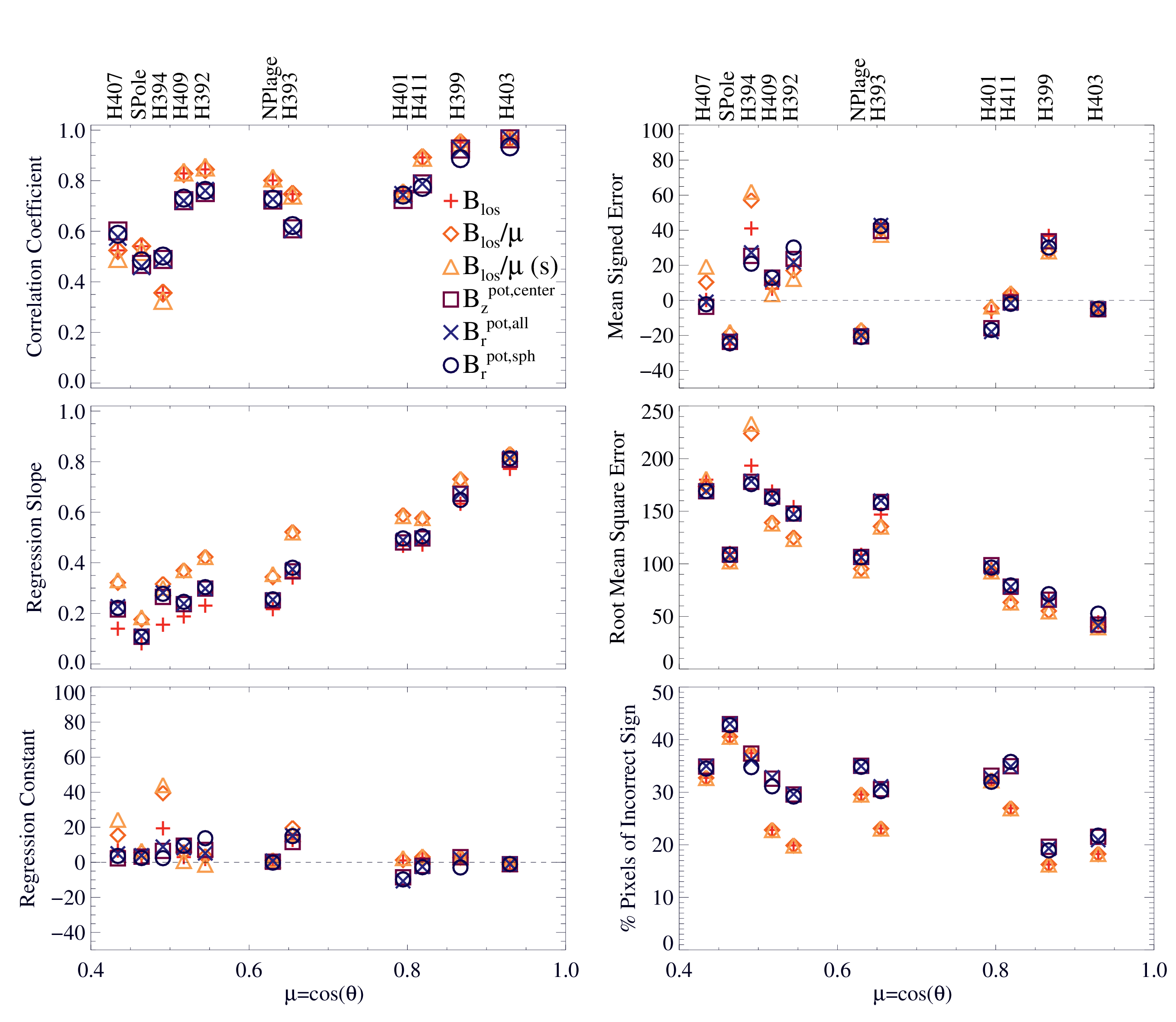}}
\caption{ Same as Figure~\ref{fig:stats_all} but only for points with
$\Bh < 500$G and $|\Br| < 500$G (note the different scales for some metrics).  Some metrics once again show strong 
relationship with $\mu$, and again there presents less of a strong trend
between the method categories.}
\label{fig:stats_weak}
\end{figure}

\begin{figure}
\centerline{
\includegraphics[width=1.00\textwidth, clip, trim = 4mm 0mm 0mm 15mm]{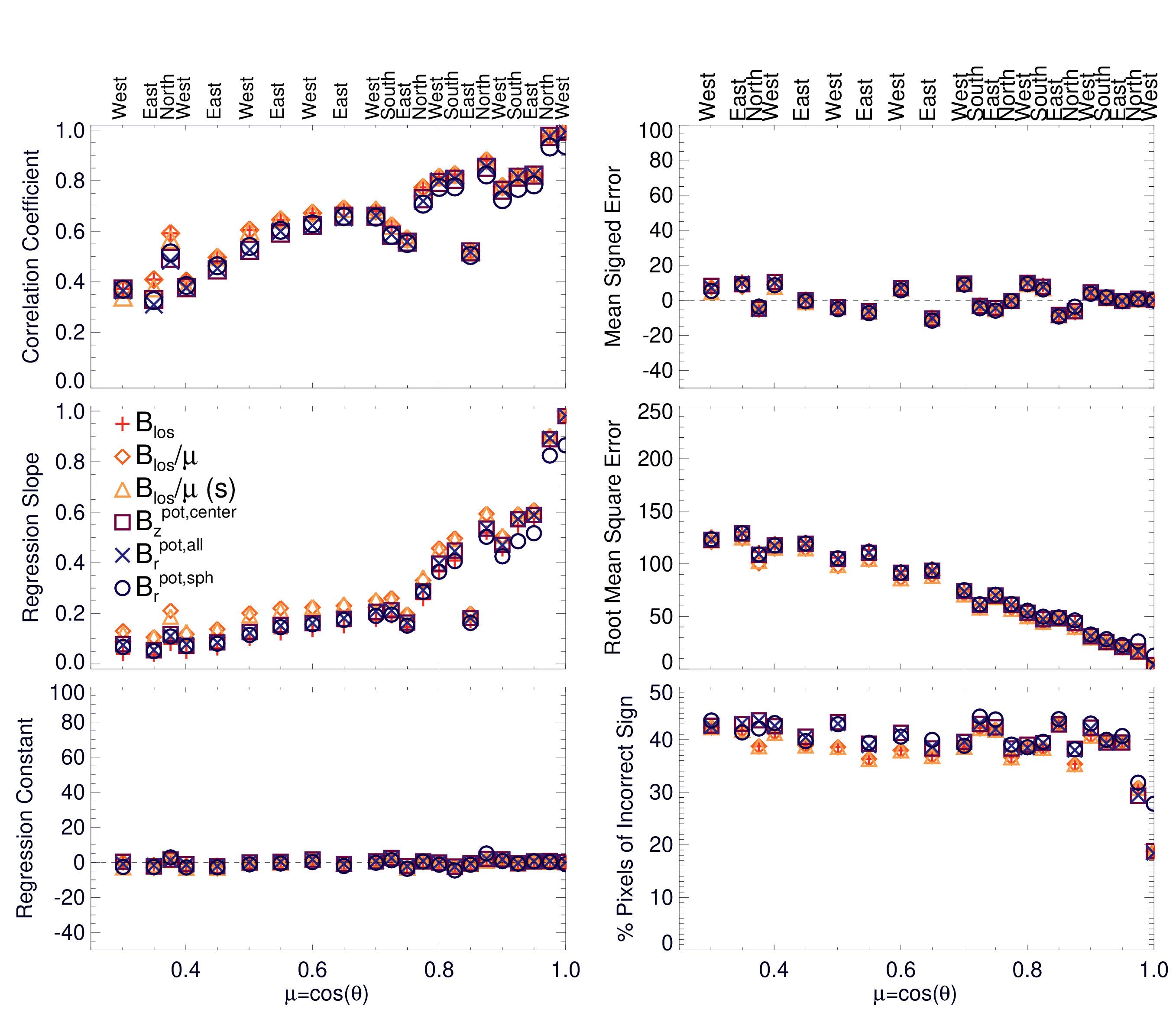}}
\caption{Same as Figure~\ref{fig:stats_all} but for the small quiet-sun
extractions at a thorough sampling of $\mu$.  Top label indicates the quadrant
from which the extraction originated.}
\label{fig:stats_qs}
\end{figure}

To summarize the performance of these approaches, 
quantitative metrics of the comparisons between $\Br$ and the different
estimations for the sub-regions on 2011.03.06 are presented graphically in
Figures~\ref{fig:stats_all},~\ref{fig:stats_strong},~\ref{fig:stats_weak},
for all HARP-based sub-regions plus the north and south targets considered, and
in Figure~\ref{fig:stats_qs} for the small quiet-sun extractions.  The metrics considered are:
the linear correlation coefficient, the fitted linear regression
slope and constant, a mean signed error, a root-mean-square
error, and the percentage of pixels that show the incorrect
sign relative to $\Br$.  All well-measured points within each
sub-region are considered in Figure~\ref{fig:stats_all} (see section ~\ref{sec:data}),
and in Figures~\ref{fig:stats_strong},~\ref{fig:stats_weak} the results are
separated between strong and weaker field areas as well.

What is clear is that there is not, in fact, a single best approach.
In some cases, {\it e.g.} for {\tt H394} and {\tt H407}, by almost
all measures the $\Brpsph$ and $\Bzppc$ approaches improve upon $\Bl$
and the $\mu$-correction methods.  The latter generally show less bias
compared to the uncorrected $\Bl$ field, but still have a large random
error. The potential-field based estimates show larger differences
for weak field strengths, but less random error.  Comparing the
weak- and strong-field results, it is clear that the small
correlations and regression slopes in the former are due to the abundance of
weak-field points and their low response to the corrections 
(see Fig.~\ref{fig:scat393} {\it vs.} \ref{fig:scatspole}). 
However, there is a general trend that plage- or weaker-field dominated areas,
including the two polar areas, are better served (under this analysis)
by the $\mu$-correction methods, notwithstanding the polarity-sign errors.
This is confirmed
as a general trend by the quiet-sun areas (Figure~\ref{fig:stats_qs}) whose
underlying structures -- like the polar area and plage areas -- are 
likely predominantly radial in HMI data.  In contrast,
when the sub-areas include or are dominated by sunspots, ({\it e.g.},
{\tt H393, H394, H401, H407}), the $\Brpsph$ and $\Bzppc$ perform the best by
these metrics.  The reasons behind this ``mixed'' message of success is
explored further, below.

Regarding the quiet-sun patches, these are sampled in order to test
the dependence of the approximations to $\mu$ only, without the complications
of different underlying structure: we assume these comprise similar
samples of small primarily radial magnetic structures.  Indeed in
Figure~\ref{fig:stats_qs}, definitive trends with $\mu$ are seen.
All models improve with increasing $\mu$ by these metrics, and there is no
obvious difference in trends between quadrants (East, North, {\it etc.}).
There are outliers, which are likely due to inherent underlying structure.
The $\mu$-correction methods generally better serve these areas, by
a small degree in some measures, than the potential-field methods.
However, all regions except those with $\mu\approx 1.0$ have a higher
percentage of points with the incorrect sign than all of the HARP regions,
except for the South Polar area.

\subsection{Total Flux Comparisons}
\label{sec:fluxcomps}

\begin{figure}
\centerline{
\includegraphics[width=0.5\textwidth, clip, trim = 8mm 10mm 0mm 5mm]{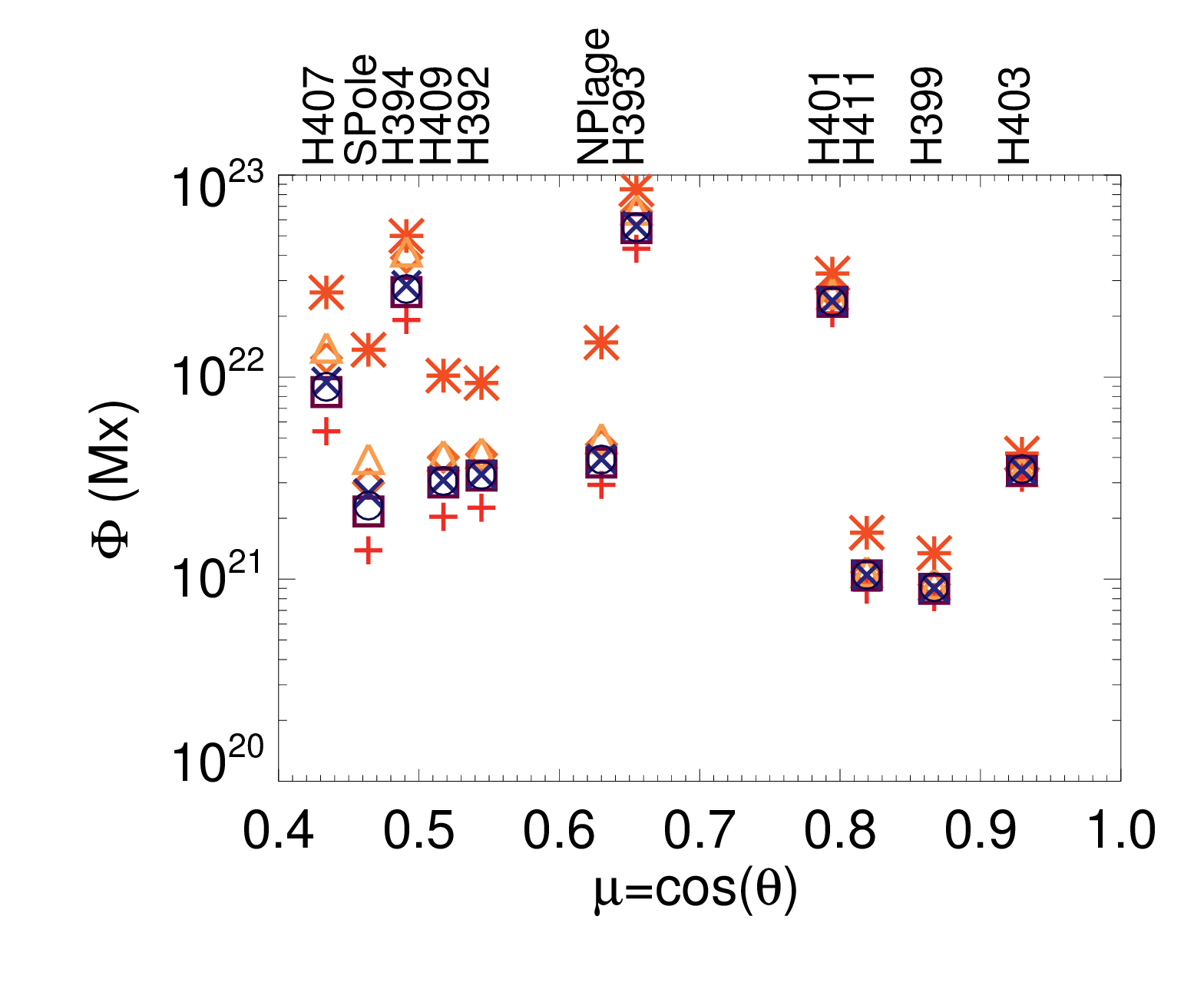}
\includegraphics[width=0.5\textwidth, clip, trim = 8mm 10mm 0mm 5mm]{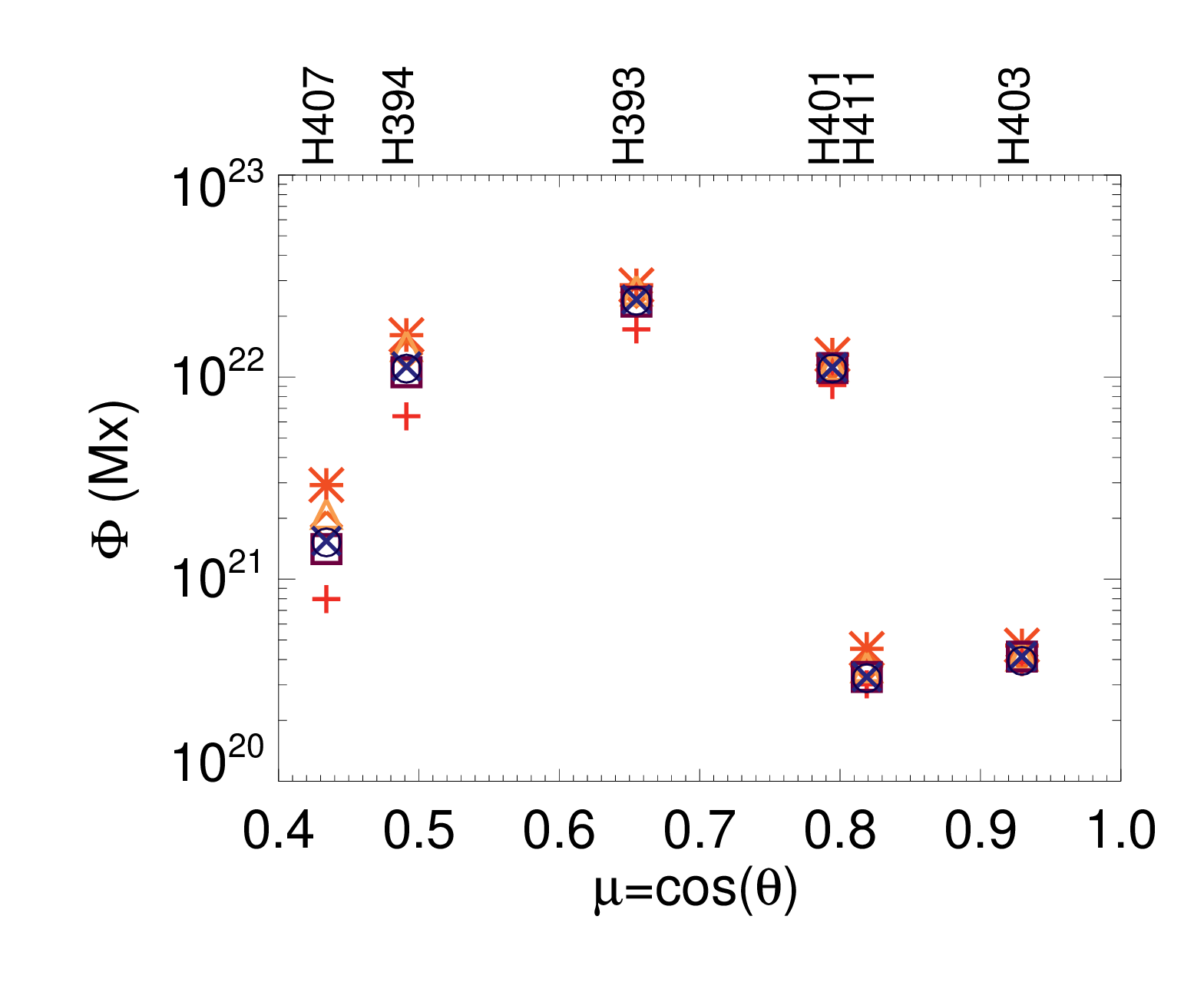}}
\caption{A comparison of inferred total magnetic flux $\Phi = \sum |{\rm B}_{\rm bndry}| dA$ 
for each sub-region,
as a function of the observing angle, using different radial-field
approximations for the $\Br$ boundary;  
for all, $dA$ is the area in Mm${}^2$ of each pixel (thus imparting some spherical
accounting for calculations which might otherwise use a planar approximation).
Left: all points in each sub-region,
Right: only those points with total field strengths over 1000\,G (which removes
some sub-regions from consideration).
For both, the symbols and colors follow Fig.~\ref{fig:stats_all} with
the addition of $\ast$: $\Br$ (red).
}
\label{fig:fluxes}
\end{figure}

The second test is the total magnetic flux of the sub-regions, where
the estimates of the magnetic flux are computed as $\sum \vert {\rm B}_{\rm
bndry} \vert d{\rm A}$ over the acceptable pixels, $dA$ is the area in Mm${}^2$
of each pixel (thus imparting some spherical accounting for the flux
which otherwise uses a planar approximation), and ${\rm B}_{\rm bndry} =
\Bl,~\Bzm,~\Bzppc$ or similar as indicated.  The results are summarized
in Figure~\ref{fig:fluxes}, both for all well-measured pixels and then
for only strong-field (sunspot) pixels.  The results for the regions
extracted from the full-disk data are considered first.

Overall, the total flux estimate using $\Phi(\Br)$ is always the
largest, that using $\Phi(\Bl)$ is always the smallest, with other approximations
in varying order between.
As demonstrated in section~\ref{sec:arcomps} the behaviors can be quite
mixed between strong-field, sunspot areas and plage areas, making the
summations over the entire sub-regions (for the total flux) difficult
to interpret.  What is also clear is that the degree of underestimation
of the $\Bl$-based flux $\Phi(\Bl)$ is a function of $\mu$, and the
flux from the $\Bzm$ boundaries do well overall at recovering the 
$\Phi(\Br)$ for the same data. The fluxes based on potential-field based
boundaries also do not completely recover $\Phi(\Br)$, even when the
area under consideration is restricted to the sunspots.  The different
implementations of each method do not vary significantly between each
other.

Further, a close examination of Figure~\ref{fig:fluxes} as compared to
Figure~\ref{fig:scat393} shows something slightly confusing: while in
Figure~\ref{fig:scat393} there are indeed regions of the distribution
where $|\Bl| > |\Br| $ and certainly $|\Bl| / \mu > |\Bl| $, in Figure~\ref{fig:fluxes}
$\Phi(\Br) > \Phi(\Bl) $ and indeed $\Phi(\Br) > \Phi(\Brms)$ always.  The differences
decrease with increasing $\mu$ as noted above.  But with a non-trivial number of
points having $|\Bl| > |\Br| $, why is the total flux consistently
larger when computed using $\Br$?  The answer is that $\Br$ includes
the higher-noise component $\Bt$, whereas any estimation using $\Bl$
does not include that higher-noise component.  The impact is much larger
in weak-signal areas which dominate the summation for total flux when
all points are used (Figure~\ref{fig:fluxes}, left) and the impact is
less when only strong-field points are included (Figure~\ref{fig:fluxes},
right).  This impact of photon noise on $\Bt$ and its particular influence
on the calculation of $\Phi(\Br)$ {\it vs.} $\Phi(\Bl)$ is confirmed using
model data with varying amount of photon noise \cite{ambigworkshop2}.
As such, while in this context we consider $\Br$ the ``answer'', it is
clear that it instead represents solely an observed estimate against
which we compare other estimates, and is likely to be an overestimate 
of the true flux.

\begin{figure}
\centerline{
\includegraphics[width=0.75\textwidth, clip, trim = 5mm 3mm 2mm 5mm]{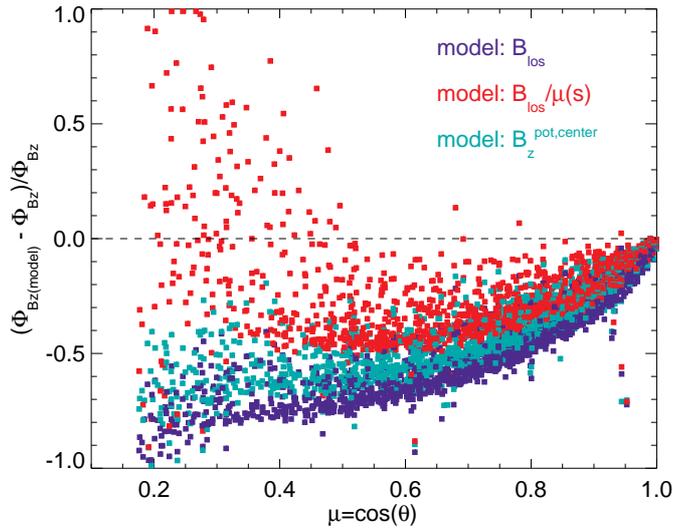}}
\caption{Scatterplot of the differences in total flux when computed using
$\Bl$, $\Brms$, and $\Bzppc$ as indicated, as a percentage difference
from the total flux computed using $\Br$ (capped at 100\% difference),
as a function of observing angle $\mu=\cos(\theta)$ (limited to $\theta
\le 80^\circ$ from disk center).  }
\label{fig:fluxes_lots}
\end{figure}

The larger HARP database is used next to examine the $\Phi({\rm
B}_{\rm bndry})$ differences for a large number of extracted regions
(Figure~\ref{fig:fluxes_lots}).   In this plot, the general underestimation
of $\Phi(\Br)$ by $\Phi(\Bl)$ is present as expected, and varies with
$\mu=\cos(\theta)$; $\Phi(\Bzppc)$ also underestimates the flux relative
to $\Phi(\Br)$ although not as severely, and the consistently larger
$\Phi(\Br)$ is now understood, from the comments above regarding the inherent
influence of noise.  $\Phi(\Brms)$
is actually closer to $\Phi(\Br)$ for much of the range in observing
angle, however it is also capable of overestimating the total flux, from
relatively modest through large observing angles; this is a property
not generally seen when using the $\Bzppc$ boundary.  
As such, using $\Phi(\Bl)$ results in the largest systematic error,
while $\Phi(\Brms)$ results in the smallest systematic error, with
$\Phi(\Bzppc)$ showing an intermediate systematic error relative to
$\Phi(\Br)$.  For the random error, the converse order holds, with
$\Phi(\Brms)$ resulting in the largest random error, $\Phi(\Bl)$ the
smallest, with the $\Phi(\Bzppc)$ random error comparable to $\Phi(\Bl)$.

\subsection{Magnetic Polarity Inversion Lines}
\label{sec:rcomps}

The introduced apparent magnetic polarity changes at the edges of
sunspots were very early signatures that inclined structures are
prevalent in active regions.  For many solar physics investigations,
however, the location of, and character of the magnetic field nearby
the magnetic polarity inversion line (PIL) is central to the analysis.
In particular, what is often of interest are magnetic PILs with locally
strong gradients in the spatial distribution of the normal field,
as an indication of localized very strong electric currents which are
associated with subsequent solar flare productivity \cite{Schrijver2007}.
Incorrect neutral lines introduced by projection may mis-identify
limb-ward penumbral areas as being strong-gradient regions of interest.

\begin{figure}
\centerline{
\includegraphics[width=0.5\textwidth, clip, trim = 16mm 10mm 22mm 2mm]{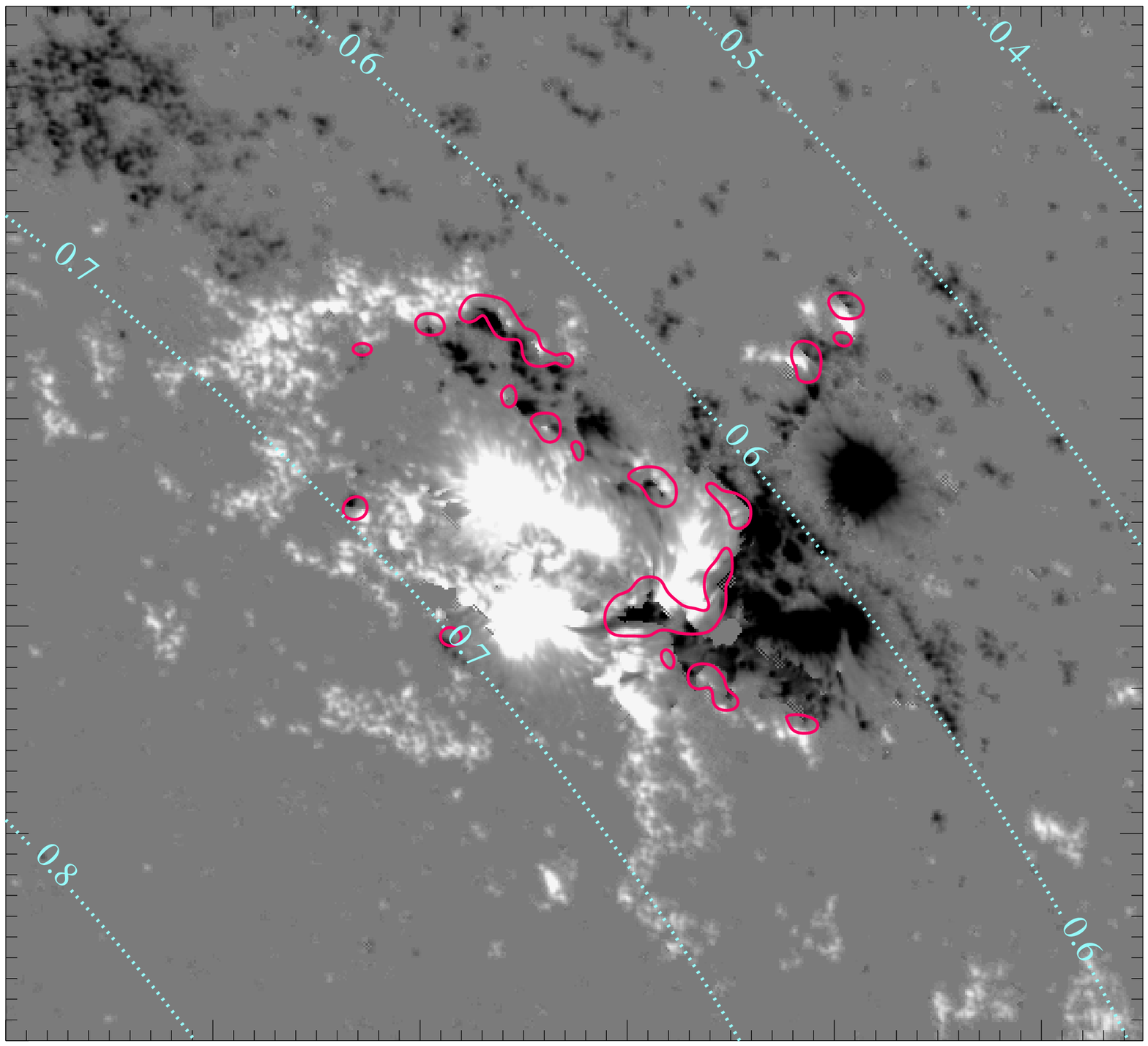}
\includegraphics[width=0.5\textwidth, clip, trim = 16mm 10mm 22mm 2mm]{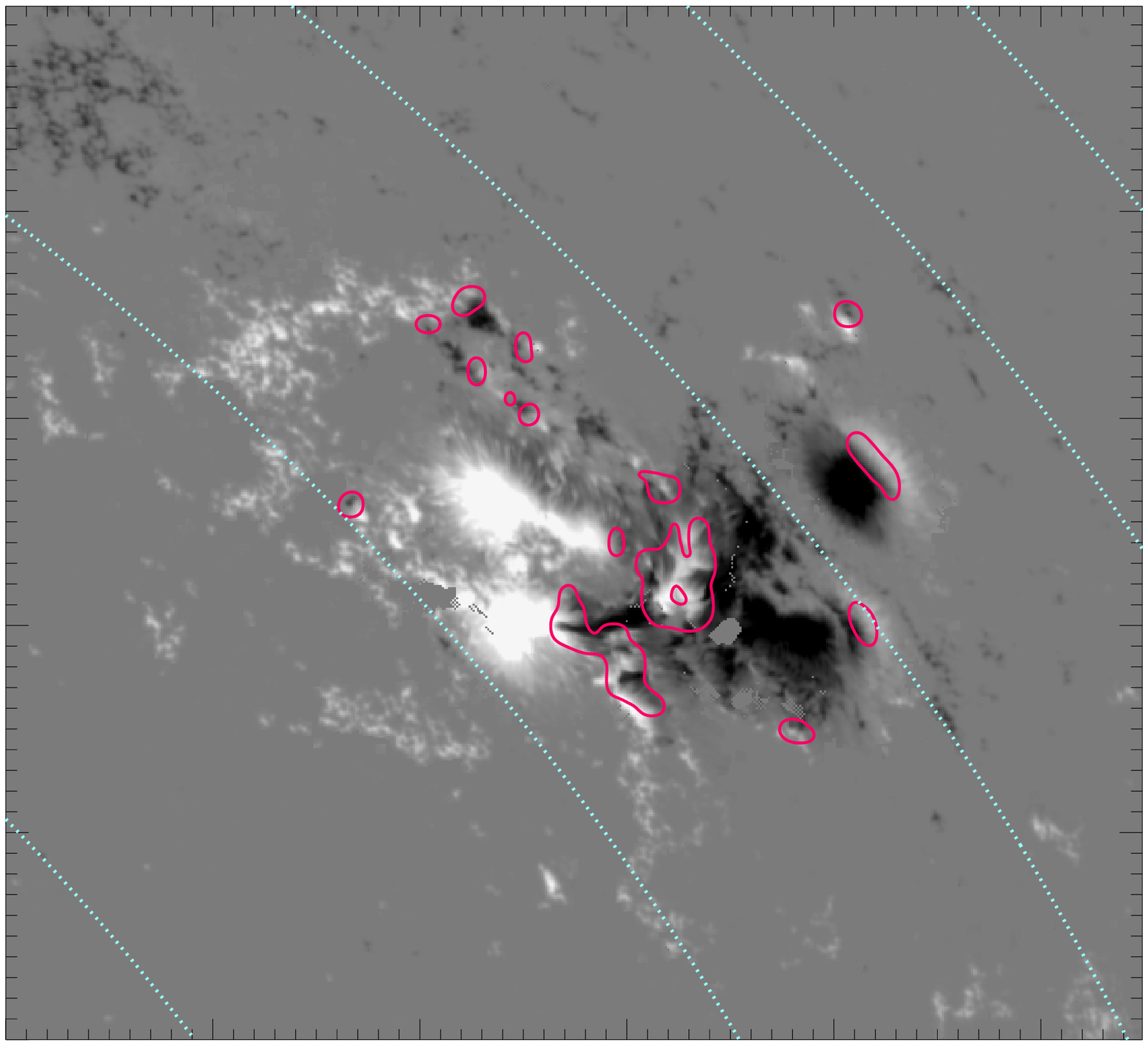}}
\centerline{
\includegraphics[width=0.5\textwidth, clip, trim = 16mm 10mm 22mm 2mm]{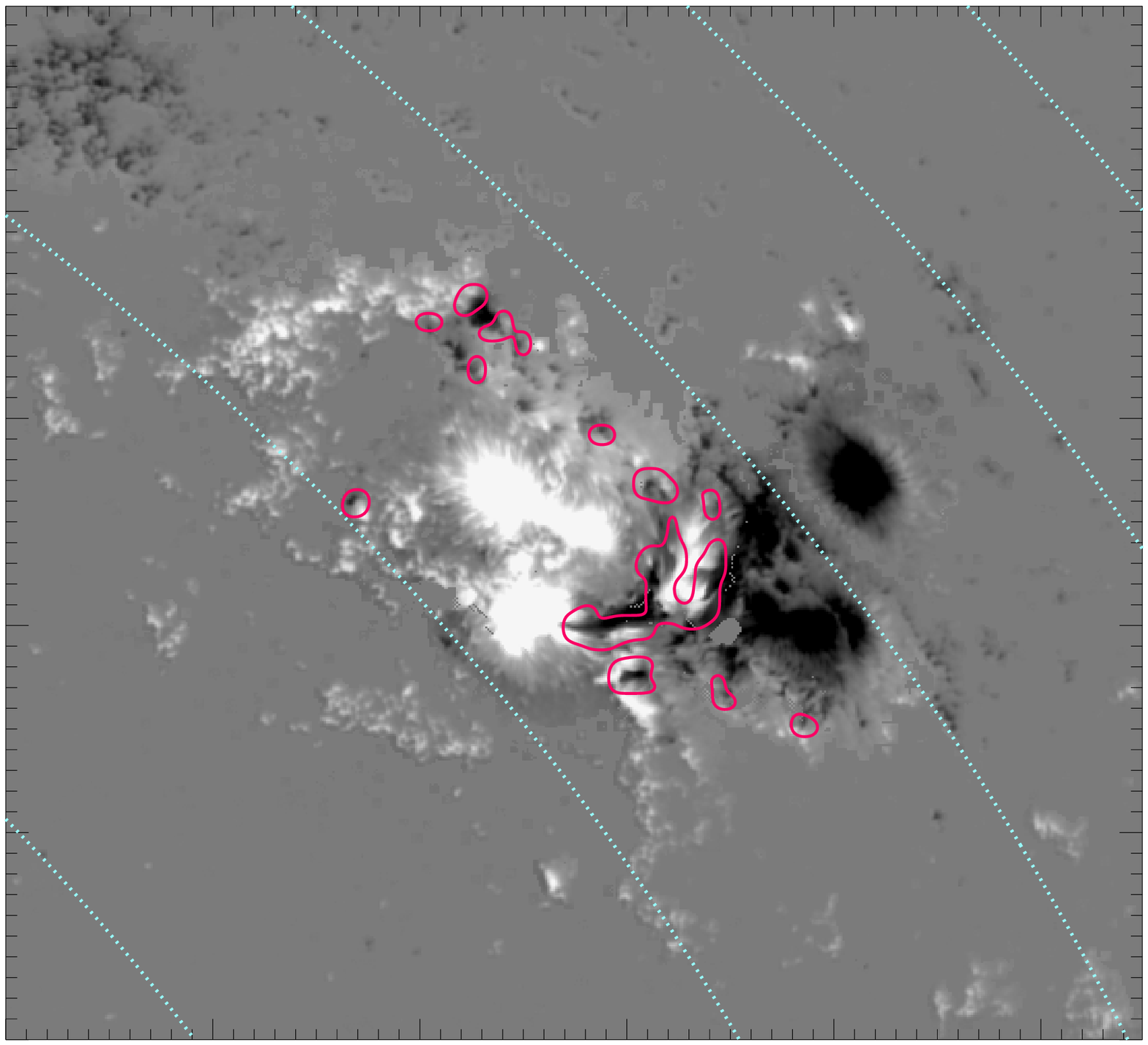}
\includegraphics[width=0.5\textwidth, clip, trim = 16mm 10mm 22mm 2mm]{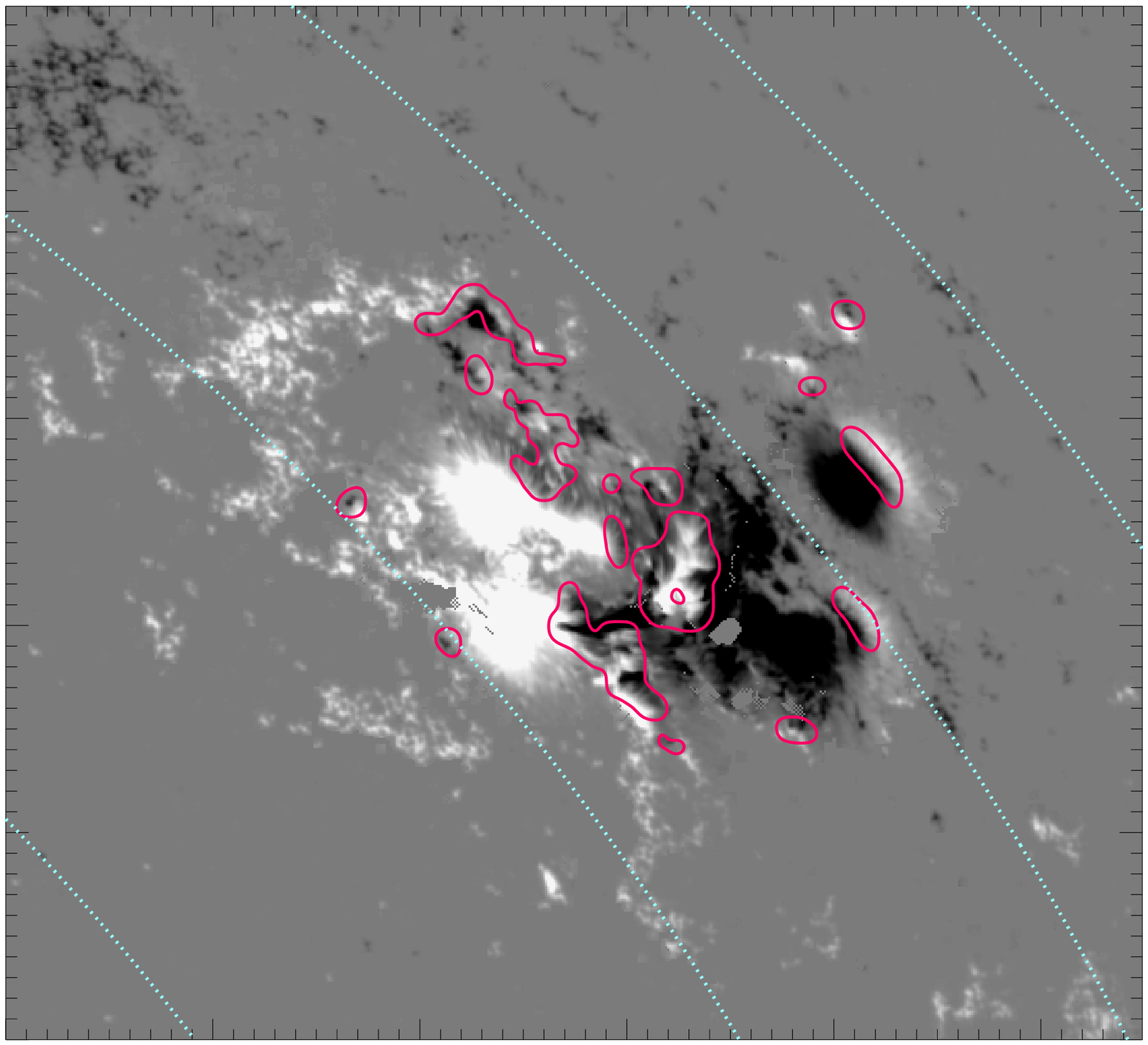}}
\caption{Images of the boundary magnetic field for NOAA AR 11164 (a
sub-area of HARP 393; {\it c.f.} Figure~\ref{fig:harps}), scaled to
$\pm1$kG, positive/negative (white/black).  The inferred strong-gradient
PILs are shown (red contours); also indicated are $\mu=\cos(\theta)$
at 0.1-spaced intervals (teal).  Top, left/right: The radial field $\Br$
(with $\mu$ contours labeled), the line-of-sight field $\Bl$.  Bottom,
left/right: The radial component of the potential field that matches
$\Bl$ on the boundary, using a planar approximation $\Bzppc$, and the
line-of-sight field with the $\mu$-correction $\Bzm$.}
\label{fig:H393rcomp}
\end{figure}

Figure~\ref{fig:H393rcomp} shows images of the observed $\Br$ field,
then the analogous estimates from $\Bl$, $\Bzm$ and $\Bzppc$ for {\tt
H393} in detail; this sunspot group is fairly large, complex, and quite
close to the limb.  In particular, areas of strong-gradient neutral lines
are highlighted.  One can see that the location of the implied magnetic
neutral line does not change at all when simply the $\Bzm$ correction
is applied to the $\Bl$ boundary, as expected from a simple scaling
factor, but the highlighted areas do change because the magnitude of the
gradient increases.  The $\Bzppc$ boundary better replicates the $\Br$
boundary, almost completely removing the introduced polarity lines on
the limb-ward sides of the sunspots.  However, it is not perfect: there
is a slight decrease in the magnitude of field in the negative-polarity
plage area which extends toward the north/east of the active region.
This is in part due to a planar approximation being invoked, but also
due to the introduction of inclined fields by the potential field model
where the underlying field inferred by HMI is predominantly vertical
(see section~\ref{sec:successNfailure}).

The recovery of a more-appropriate PIL in or near sunspots
with polarity errors in weaker fields was seen earlier
in the analysis of the ``\% Points of Incorrect Sign'' in
Figs.~\ref{fig:stats_all}--\ref{fig:stats_weak}.  The strong-field
regions showed a significant decrease in incorrect-polarity fields when a
potential-field-based boundary was used relative to the $\mu$ correction
boundaries, but in the weak field areas the results were mixed, leading
to a similarly mixed result when all pixels were included.

\subsection{Analysis of Success and Failure}
\label{sec:successNfailure}

While it is clear that the location of the magnetic neutral line 
is better recovered for sunspot areas away from disk center using a potential-field
model than
is possible using the $\Bl$ boundary and a multiplicative factor,
it is also clear that there are indeed some solar structures
for which the potential-field model does not perform well.  

We investigate where the different approximations work well, and 
where they do not, based on the hypothesis that the $\mu$-correction
approach should be exactly correct (by construction) for truly radial 
fields as inferred within the limitations of the instrument in question. 

\begin{figure}
\centerline{
\includegraphics[width=0.5\textwidth, clip, trim = 0mm 0mm 0mm 0mm]{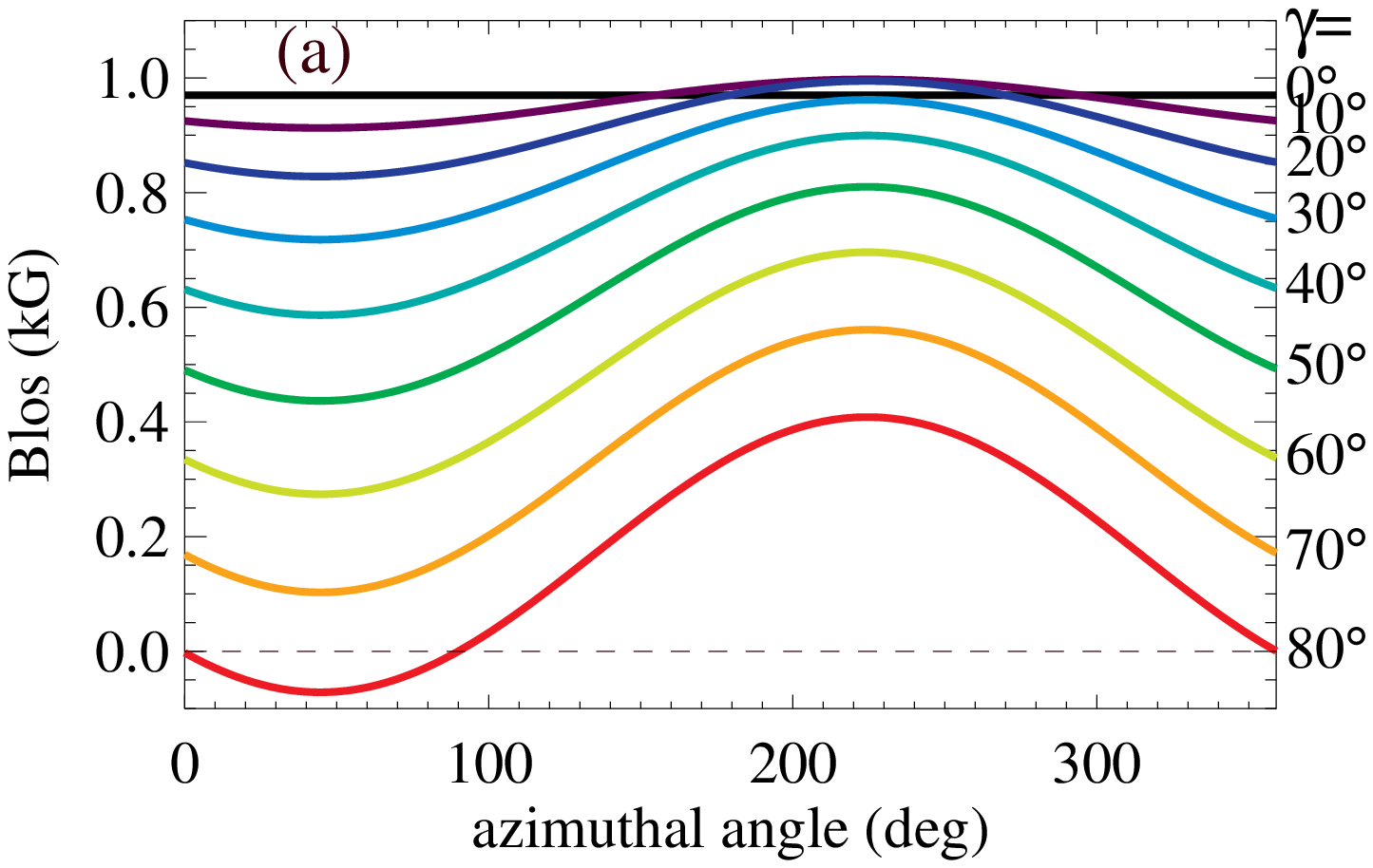}
\includegraphics[width=0.5\textwidth, clip, trim = 0mm 0mm 0mm 0mm]{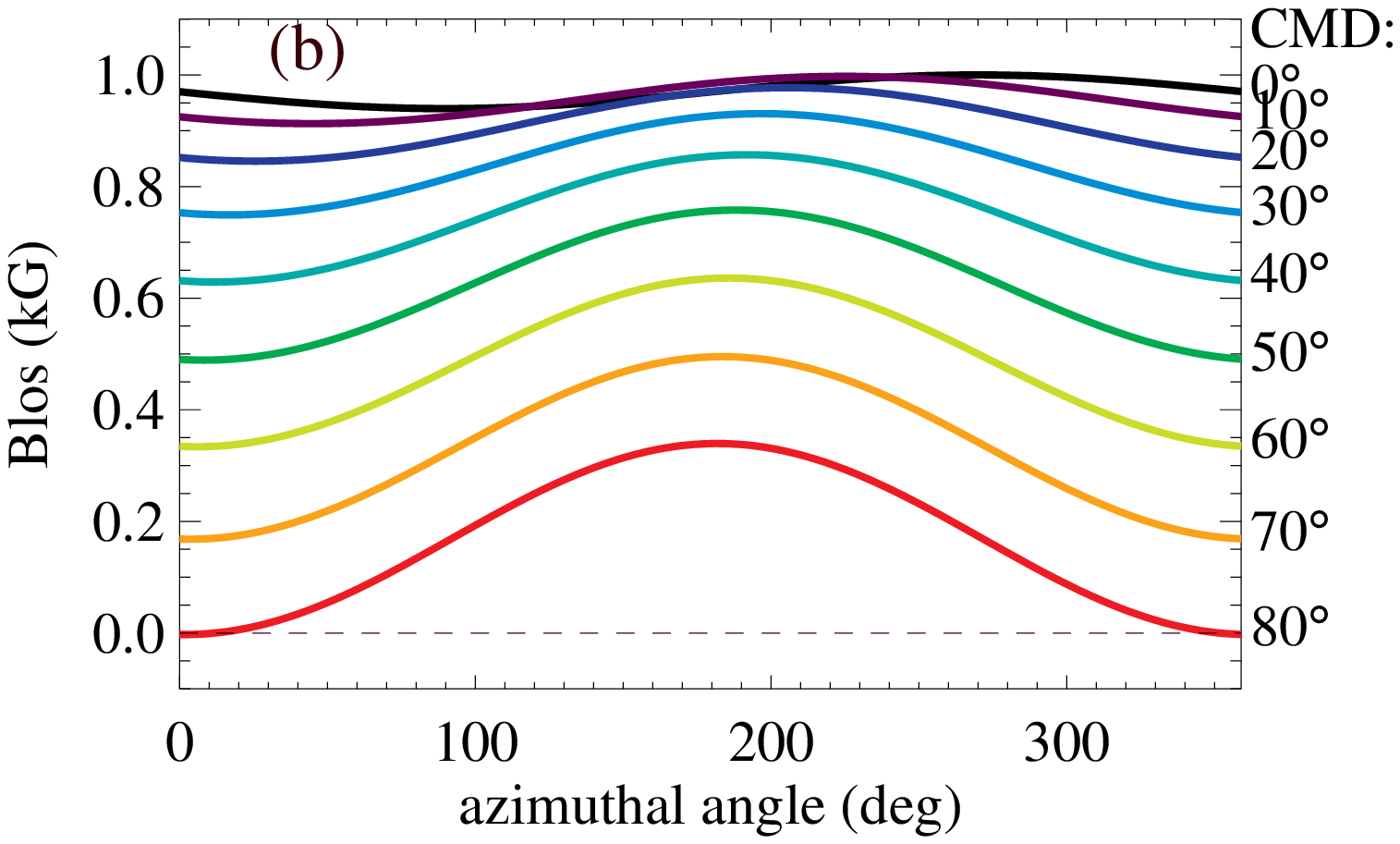}}
\caption{Variations in expected $\Bl$ signal from a magnetic vector with
$|\BB|=1000$G, located at N10, as a function of inherent azimuthal angle $\phi$,
while varying {\bf (a)} the inclination angle $\gamma$ (keeping the
heliographic longitude at W10), and {\bf (b)} varying the heliographic
longitude and thus the observing angle (keeping the inherent
inclination angle at $10^\circ$. }
\label{fig:bl_demo}
\end{figure}

First, it behooves us to recall that the $\Bl$ component of a magnetic
field vector itself can vary widely for a given magnitude $|\BB|$ as a
function of the observing angle $\theta$, the {\it local} inclination
angle $\gamma$ (relative to the {\it local} normal), and the local azimuthal
angle $\phi$.  Conventional practices of assuming that errors are
within acceptable limits when a target $\Bl$ is within, say, $30^\circ$
of disk center, may be surprisingly misleading when the expected $\Bl$
magnitude is so significantly different from the input vector magnitude,
as demonstrated in Figure~\ref{fig:bl_demo}.  The impact of the azimuthal
angle as well as the inclination angle on the observed $\Bl$ explains
some of why the $\mu$-correction based estimates of the radial field may
impart greater errors than might be expected.  In other words, one should
expect (for example) a 20\% introduced uncertainty in the $\Bl$ component
relative to the inherent field strengths even at $\theta=30^\circ$ simply
due to the unknown azimuthal directions of the underlying horizontal
component of the field -- as contrasted to an estimated 13\% error from
simply geometric considerations at this observing angle.

\begin{figure}
\centerline{
\includegraphics[width=0.95\textwidth, clip, trim = 25mm 15mm 30mm 10mm]{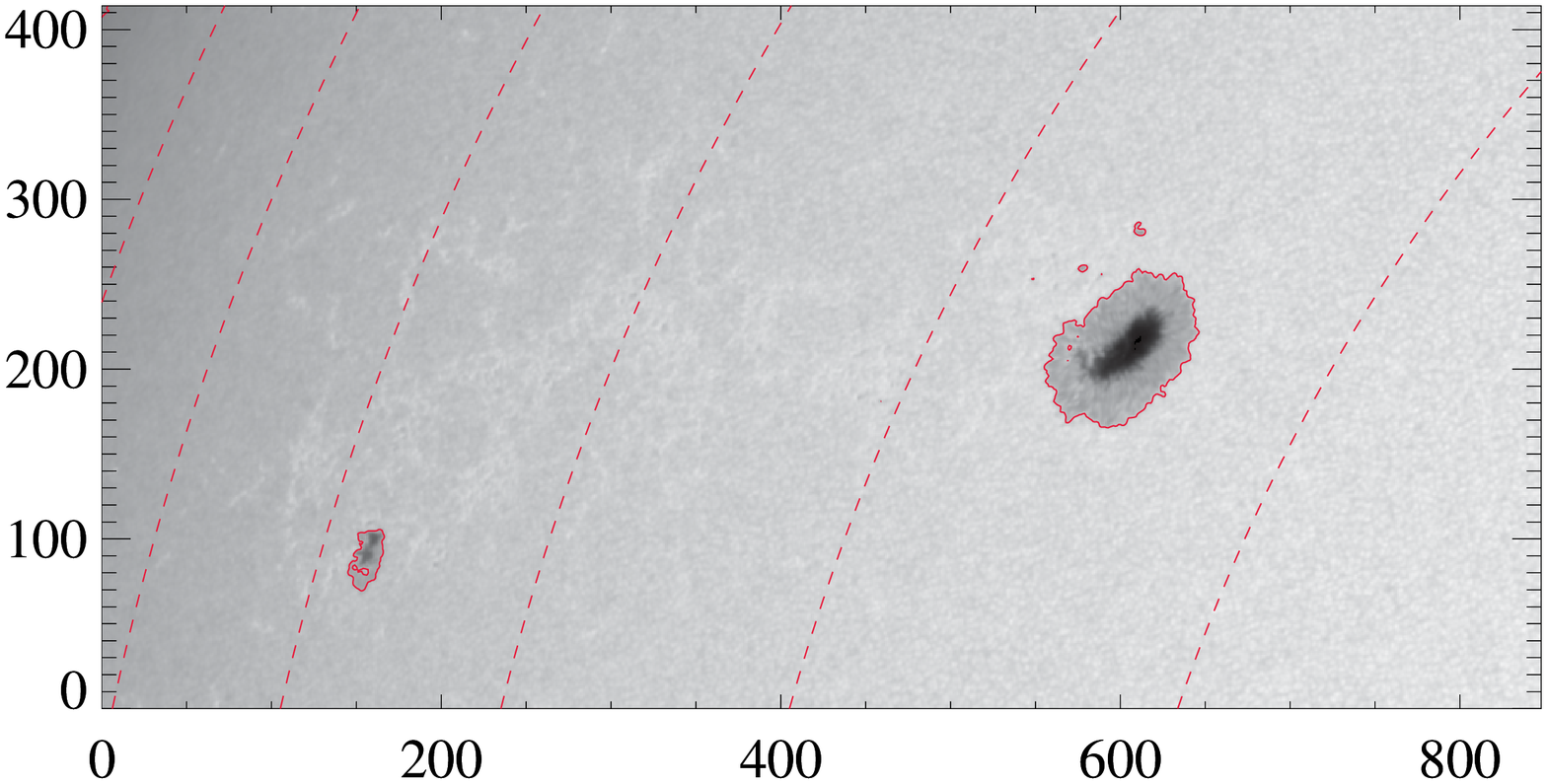}}
\centerline{
\includegraphics[width=0.95\textwidth, clip, trim = 25mm 15mm 30mm 10mm]{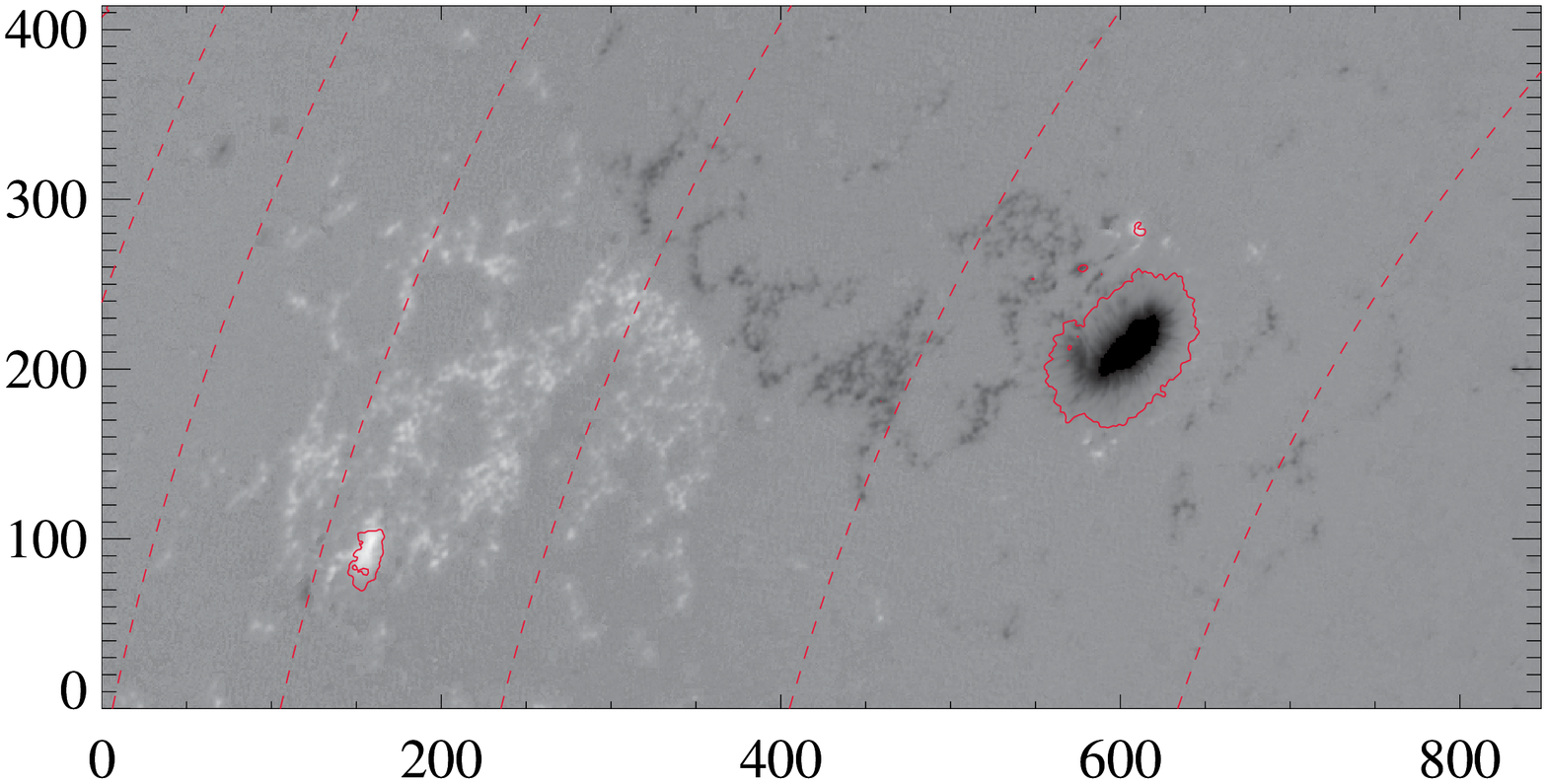}}
\caption{HARP 3848 (NOAA AR 12005, 12007), obtained at 2014.03.15 at 15:48:00 TAI.  
Top: continuum intensity, Bottom: radial field $\Bz$, 
positive/negative polarity as white/black and scaled to $\pm2$kG; red contours 
indicate 0.9 times the median of the continuum intensity and indicate the sunspots,
red dashed contours delineate $\mu= 0.3,0.4,0.5,0.6,0.7,0.8]$; for reference to 
Figure~\ref{fig:H3848diffs}.}
\label{fig:H3848context}
\end{figure}

\begin{figure}
\centerline{
\includegraphics[width=0.5\textwidth, clip, trim = 5mm 0mm 0mm 10mm]{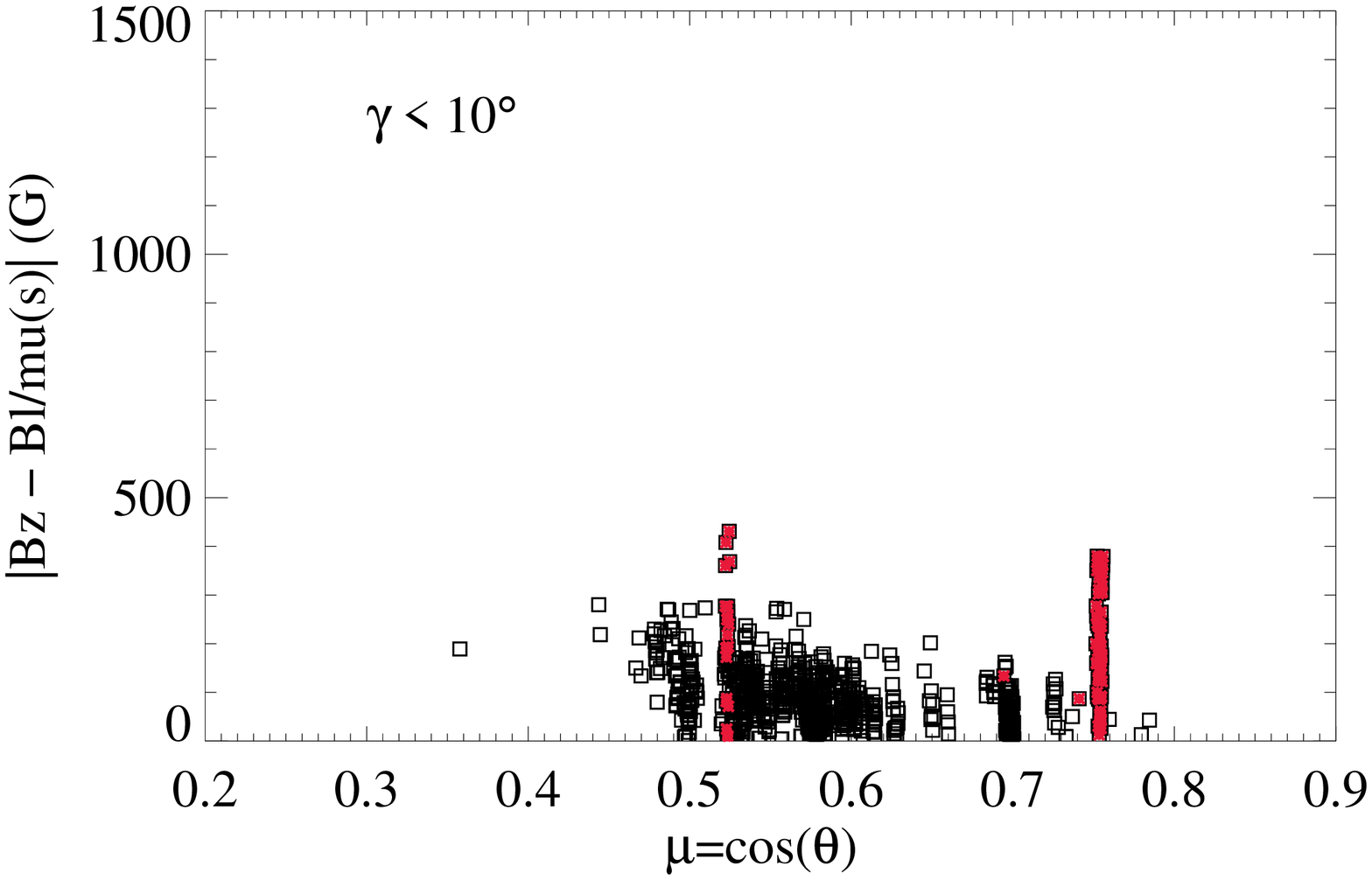}
\includegraphics[width=0.5\textwidth, clip, trim = 5mm 0mm 0mm 10mm]{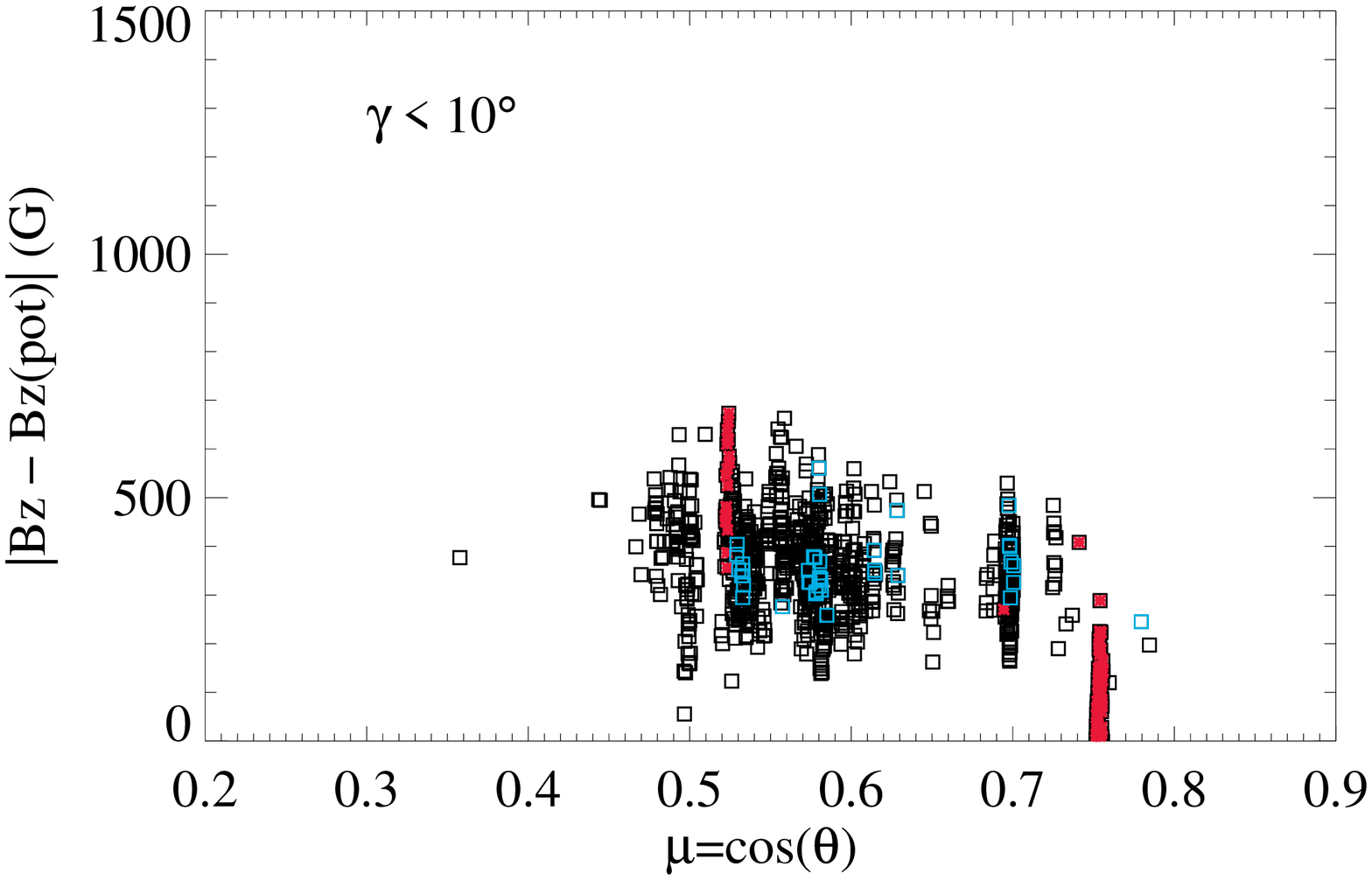}}
\centerline{
\includegraphics[width=0.5\textwidth, clip, trim = 5mm 0mm 0mm 10mm]{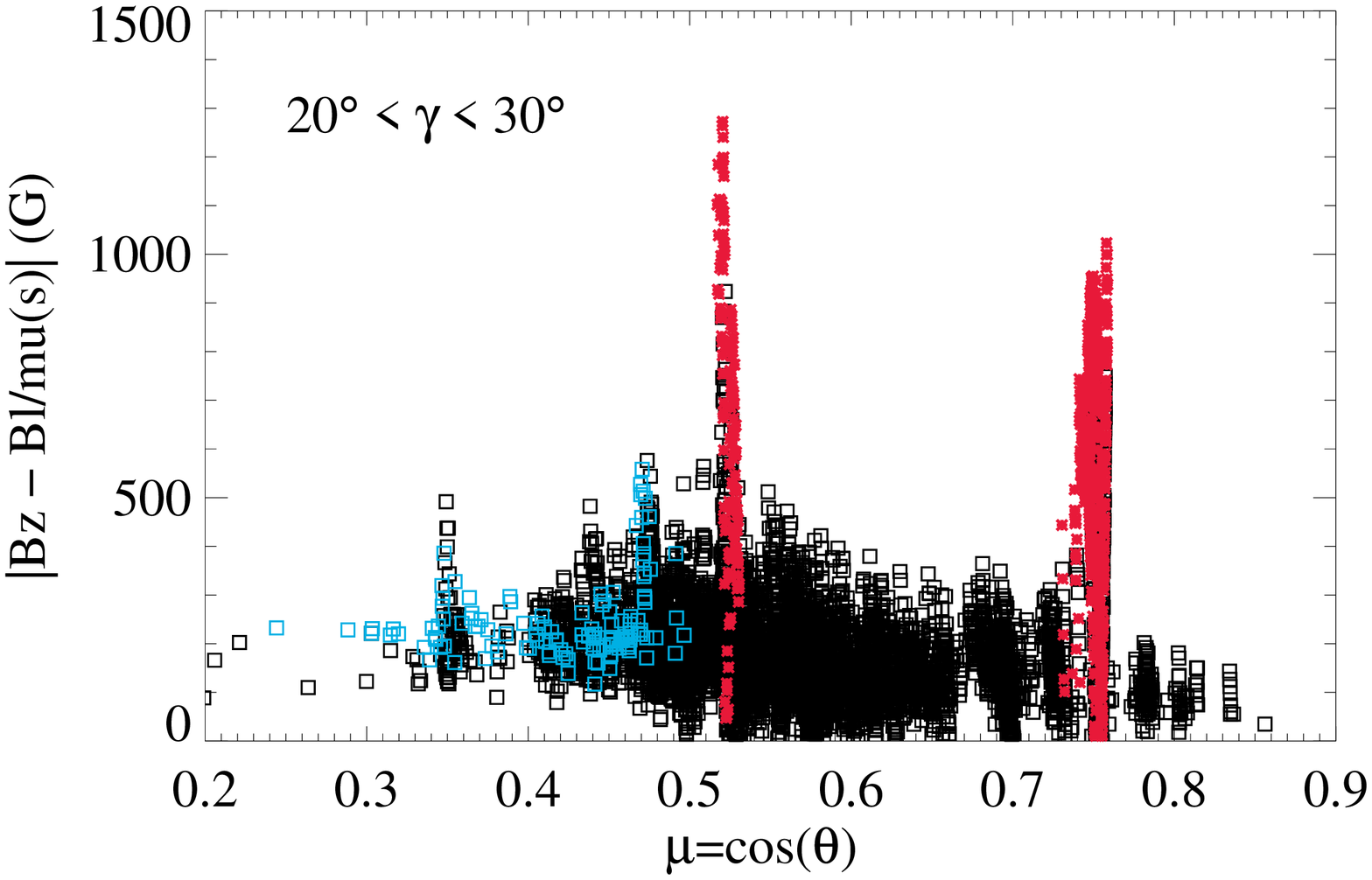}
\includegraphics[width=0.5\textwidth, clip, trim = 5mm 0mm 0mm 10mm]{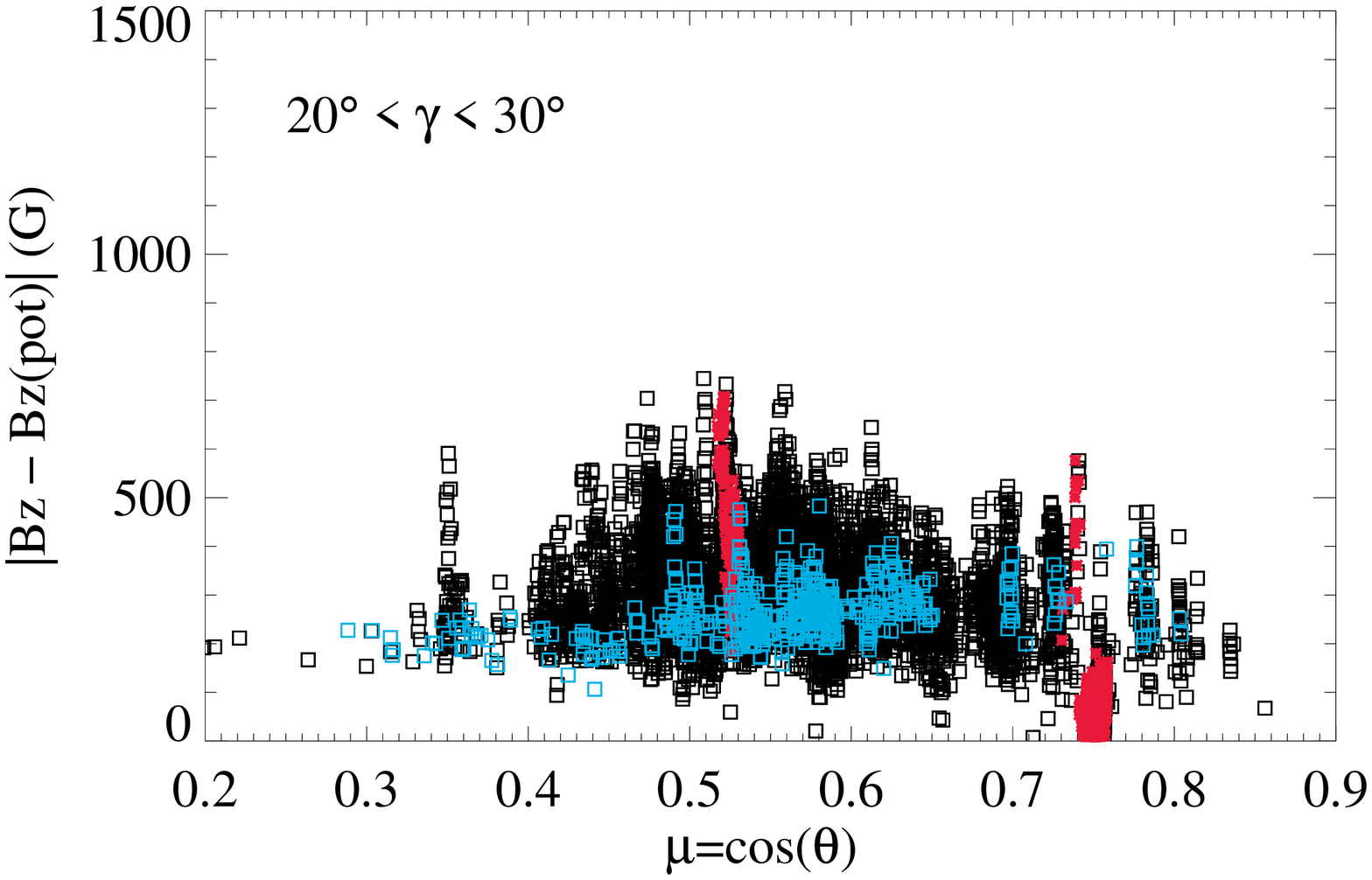}}
\centerline{
\includegraphics[width=0.5\textwidth, clip, trim = 5mm 0mm 0mm 10mm]{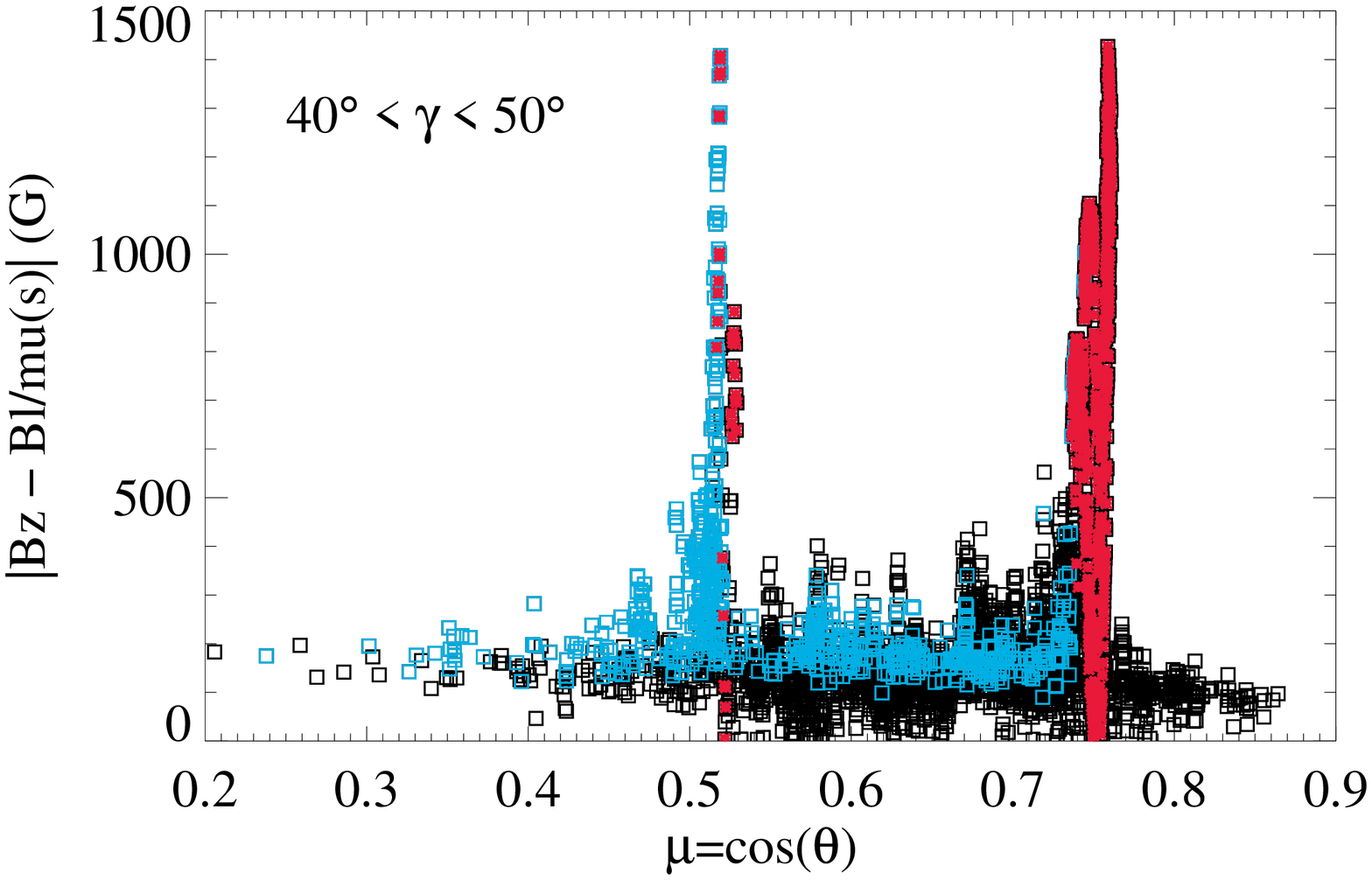}
\includegraphics[width=0.5\textwidth, clip, trim = 5mm 0mm 0mm 10mm]{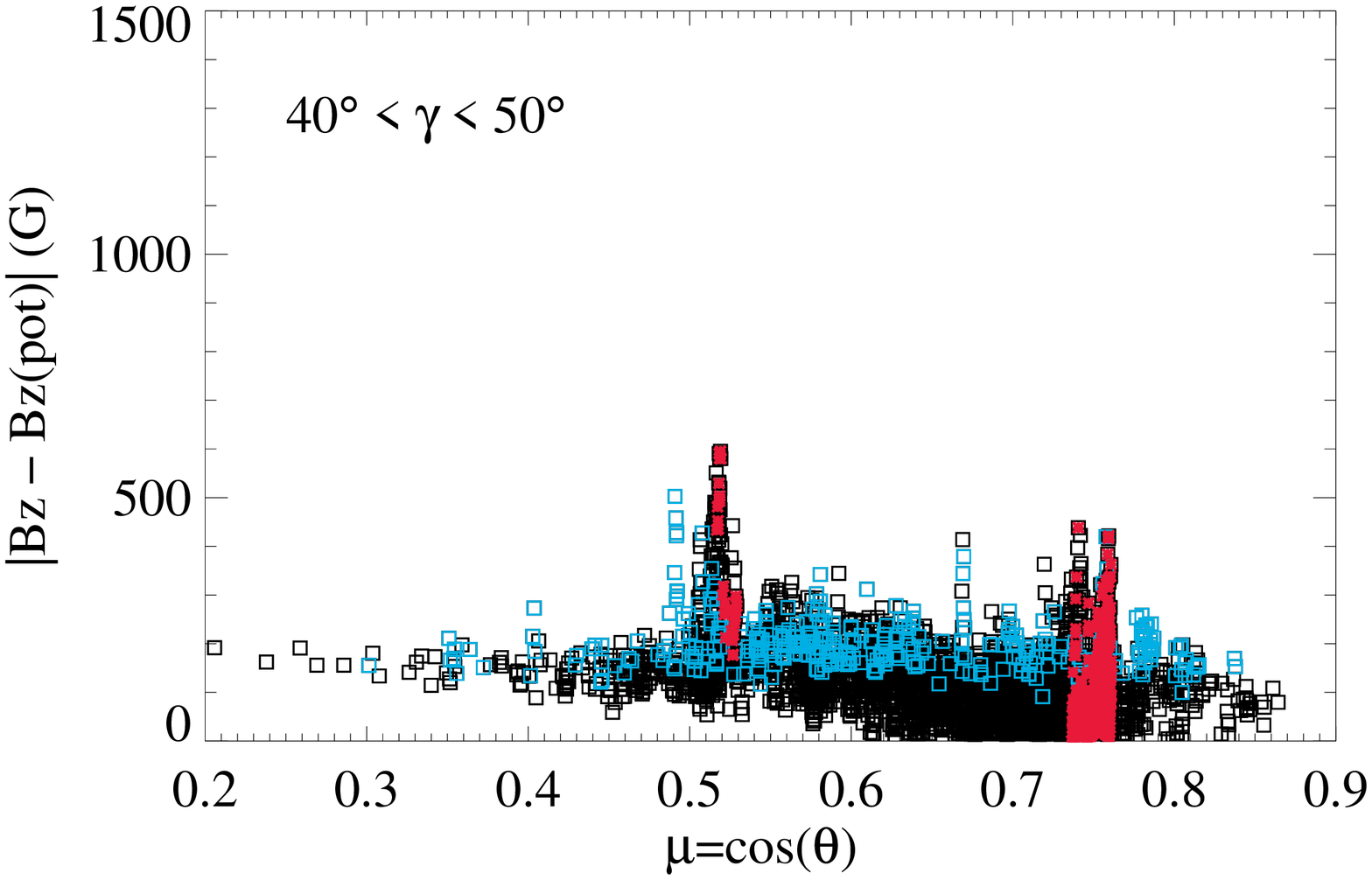}}
\caption{Top: absolute difference between $\Bz$ and $\Brms $(left), and $\Bzppc$ (right), 
as a function of $\mu=\cos(\theta)$, only for points with local inclination angle $\gamma$ 
less than $10^\circ$ from radial.  Points that lie within the spots are indicated by $\ast$ (red).
Pixels for which the model resulted in a different sign of the field are further
highlighted by over-drawing the squares in blue.
Middle, same but for points $20^\circ<\gamma<30^\circ$; Bottom: same, but for points with 
$40^\circ<\gamma<50^\circ$.}
\label{fig:H3848diffs}
\end{figure}

\begin{figure}
\centerline{
\includegraphics[width=0.5\textwidth, clip, trim = 5mm 0mm 0mm 10mm]{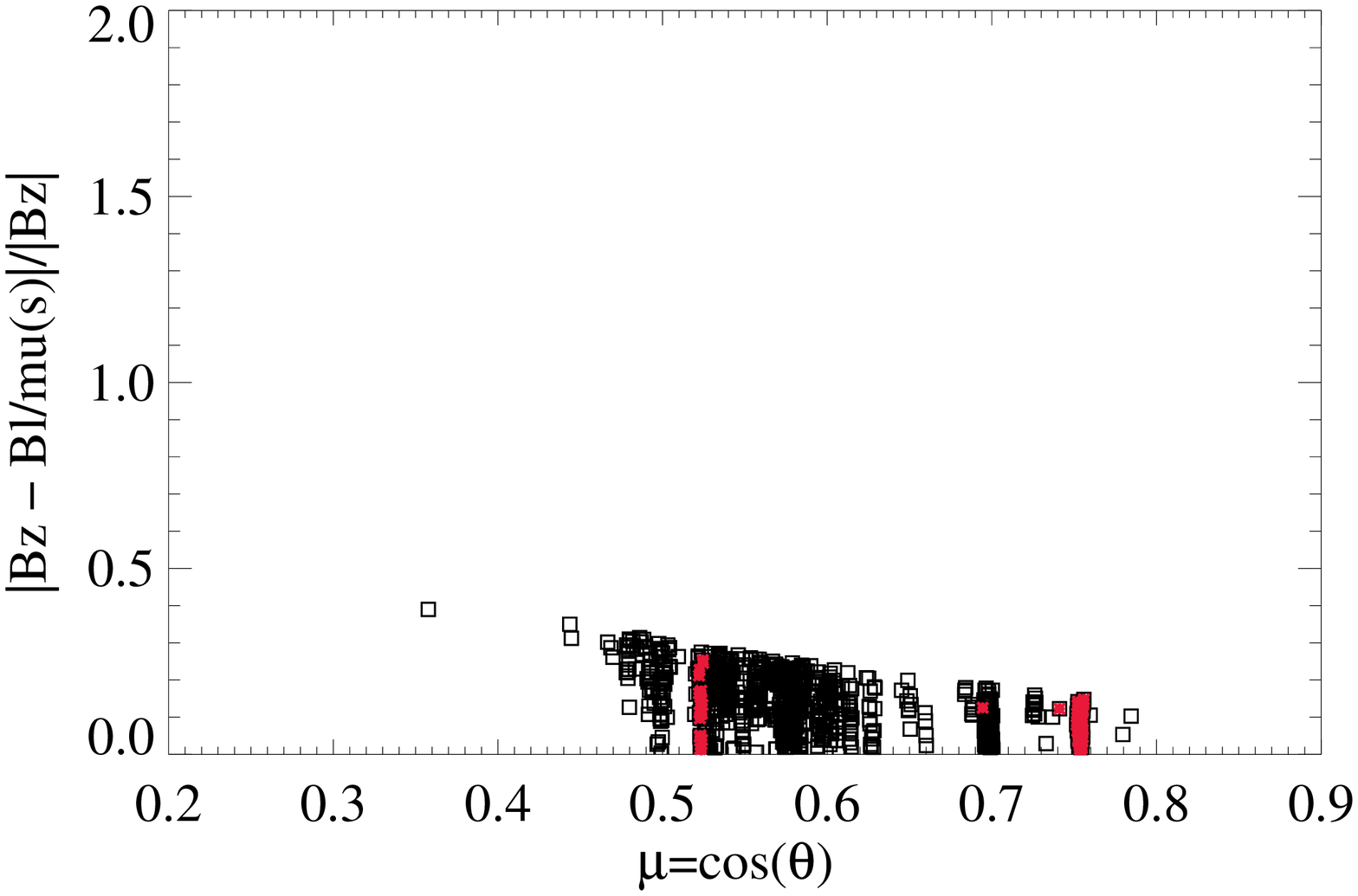}
\includegraphics[width=0.5\textwidth, clip, trim = 5mm 0mm 0mm 10mm]{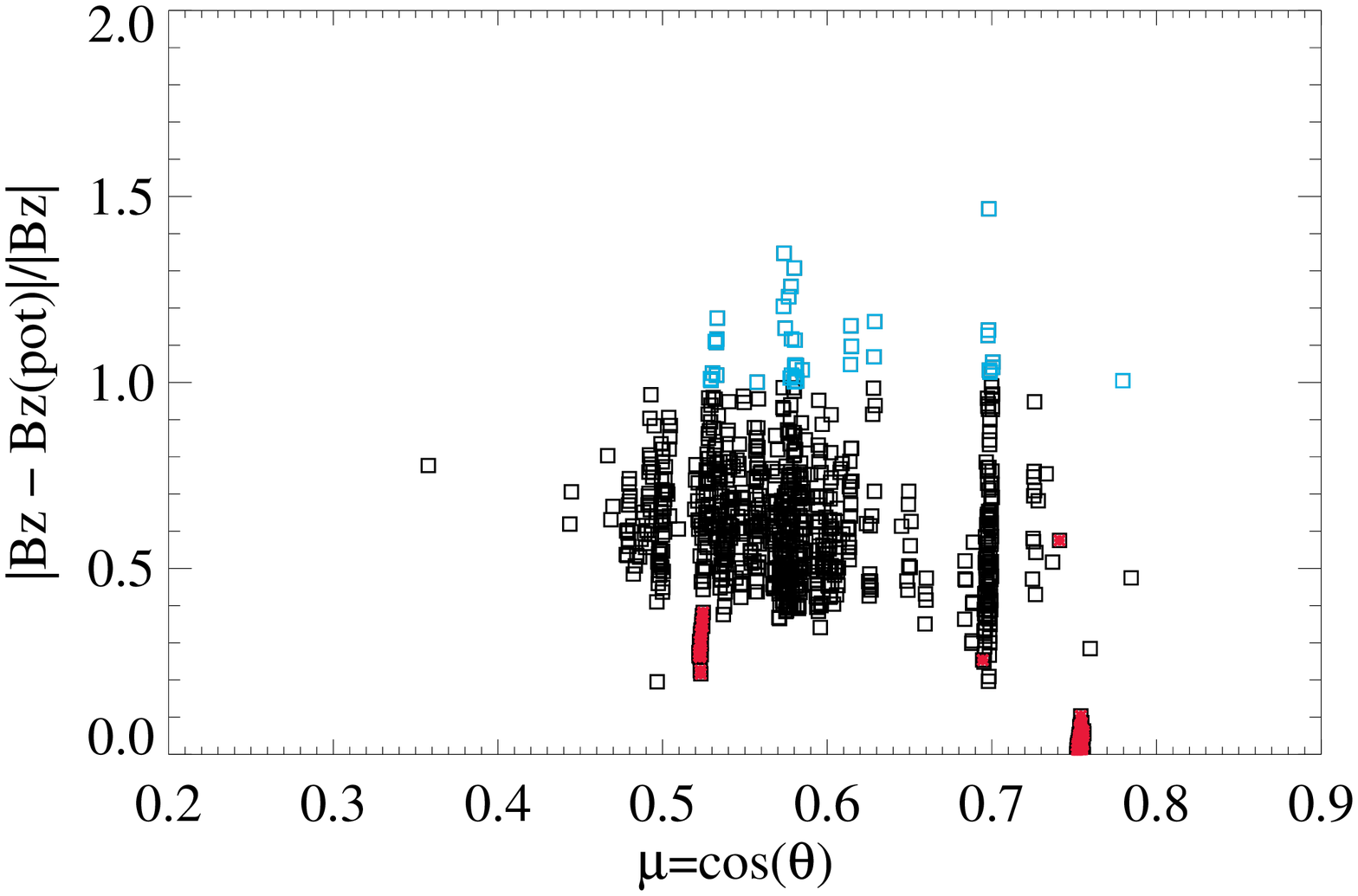}}
\centerline{
\includegraphics[width=0.5\textwidth, clip, trim = 5mm 0mm 0mm 10mm]{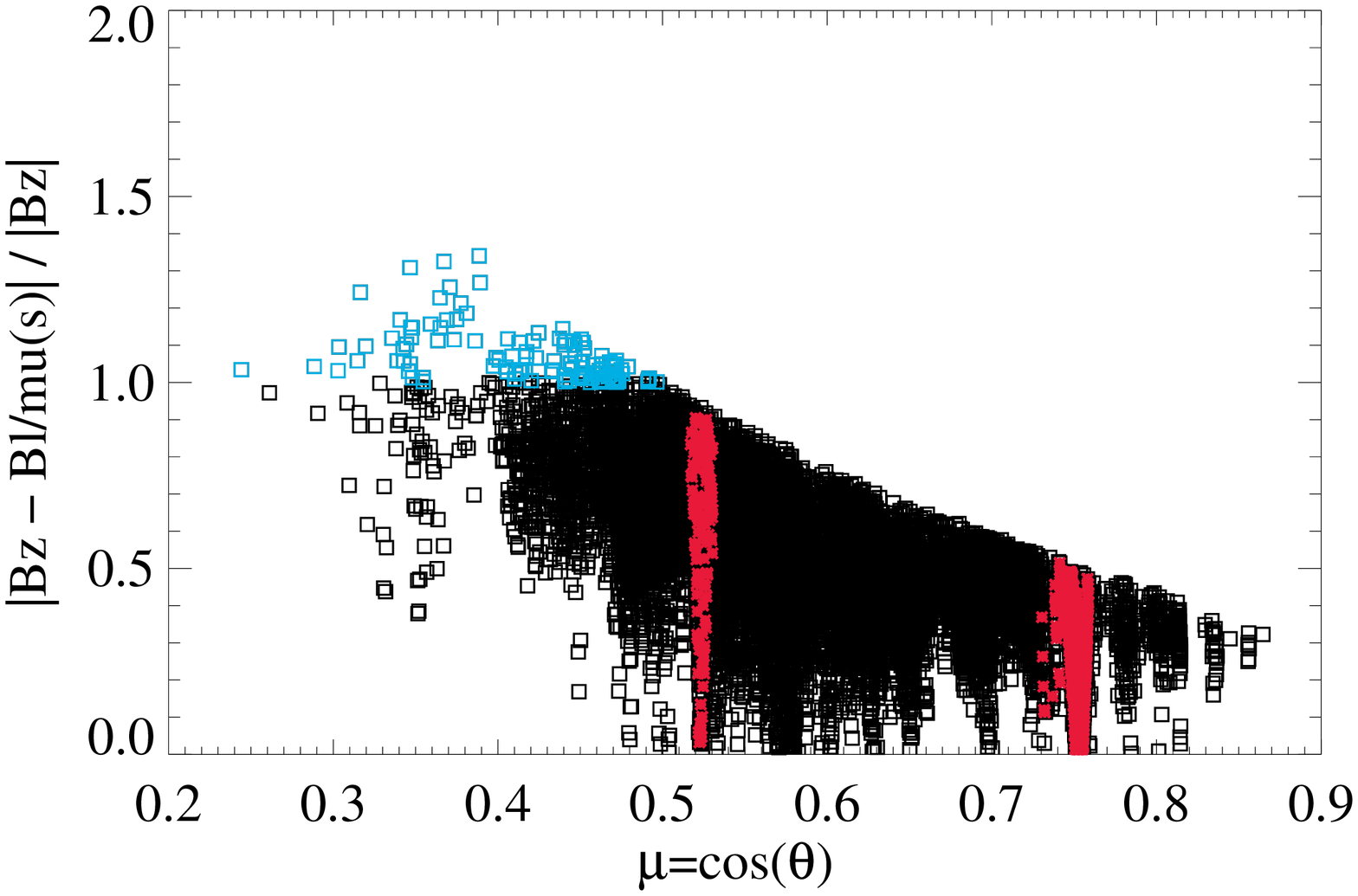}
\includegraphics[width=0.5\textwidth, clip, trim = 5mm 0mm 0mm 10mm]{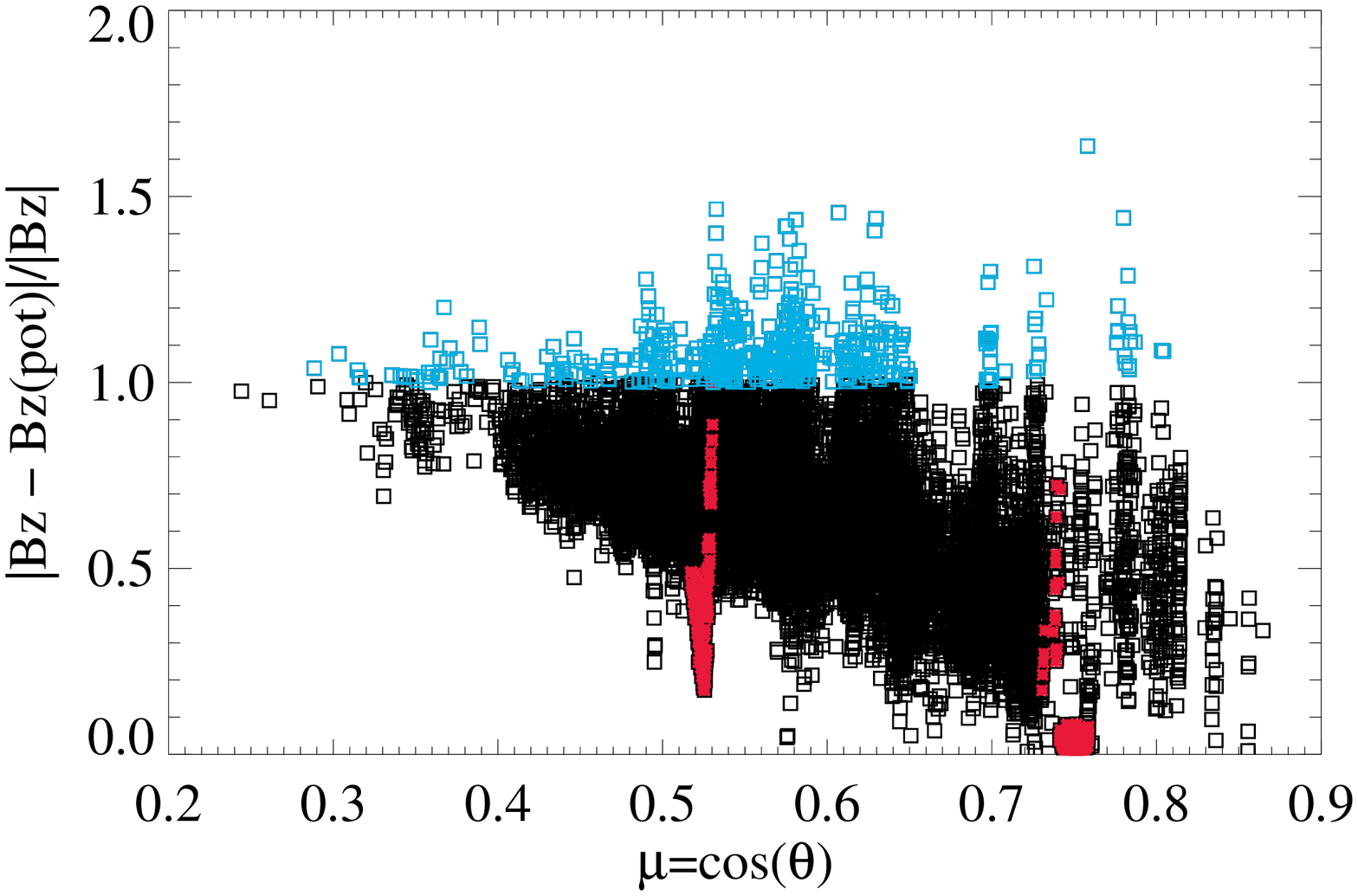}}
\centerline{
\includegraphics[width=0.5\textwidth, clip, trim = 5mm 0mm 0mm 10mm]{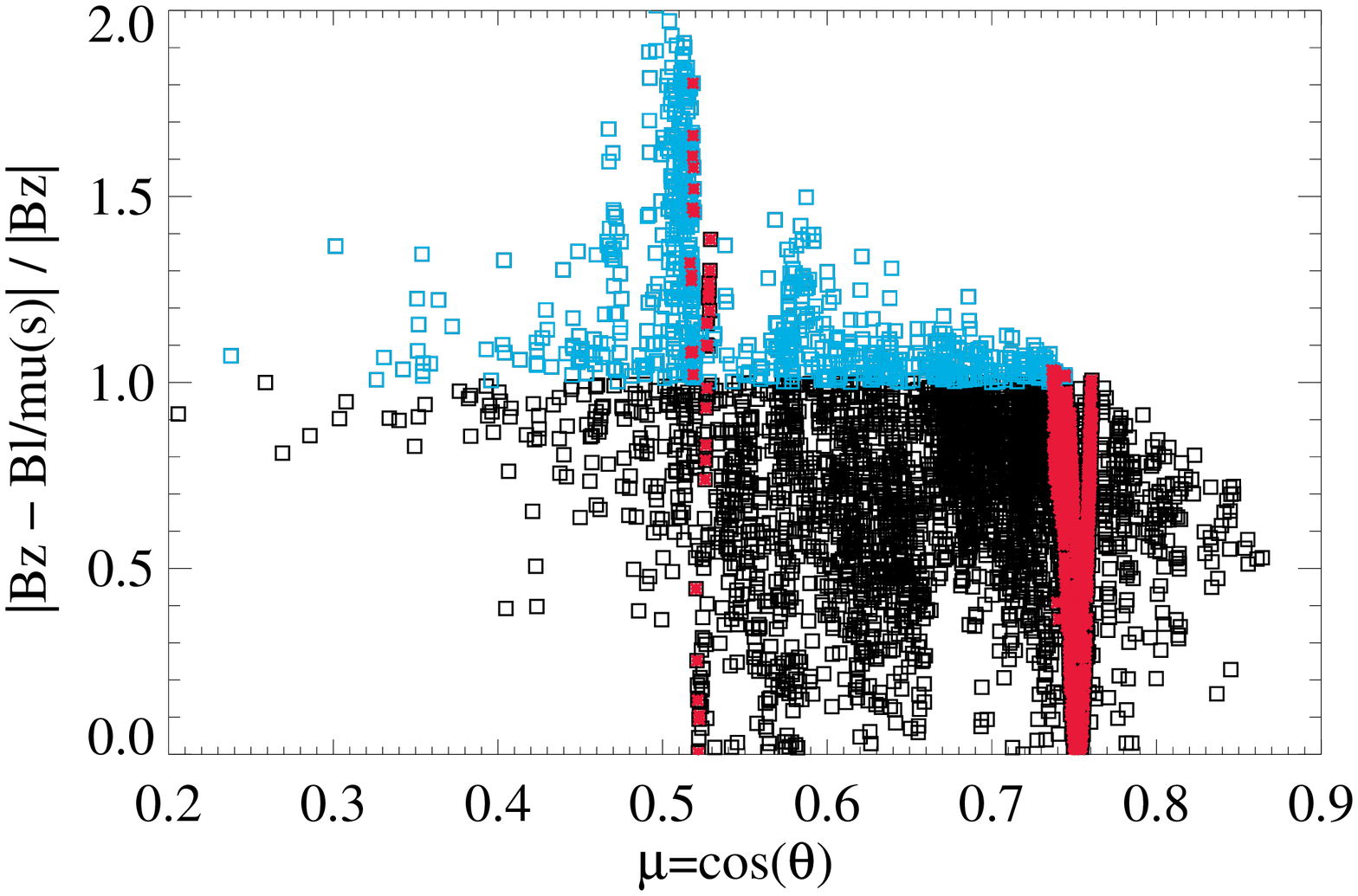}
\includegraphics[width=0.5\textwidth, clip, trim = 5mm 0mm 0mm 10mm]{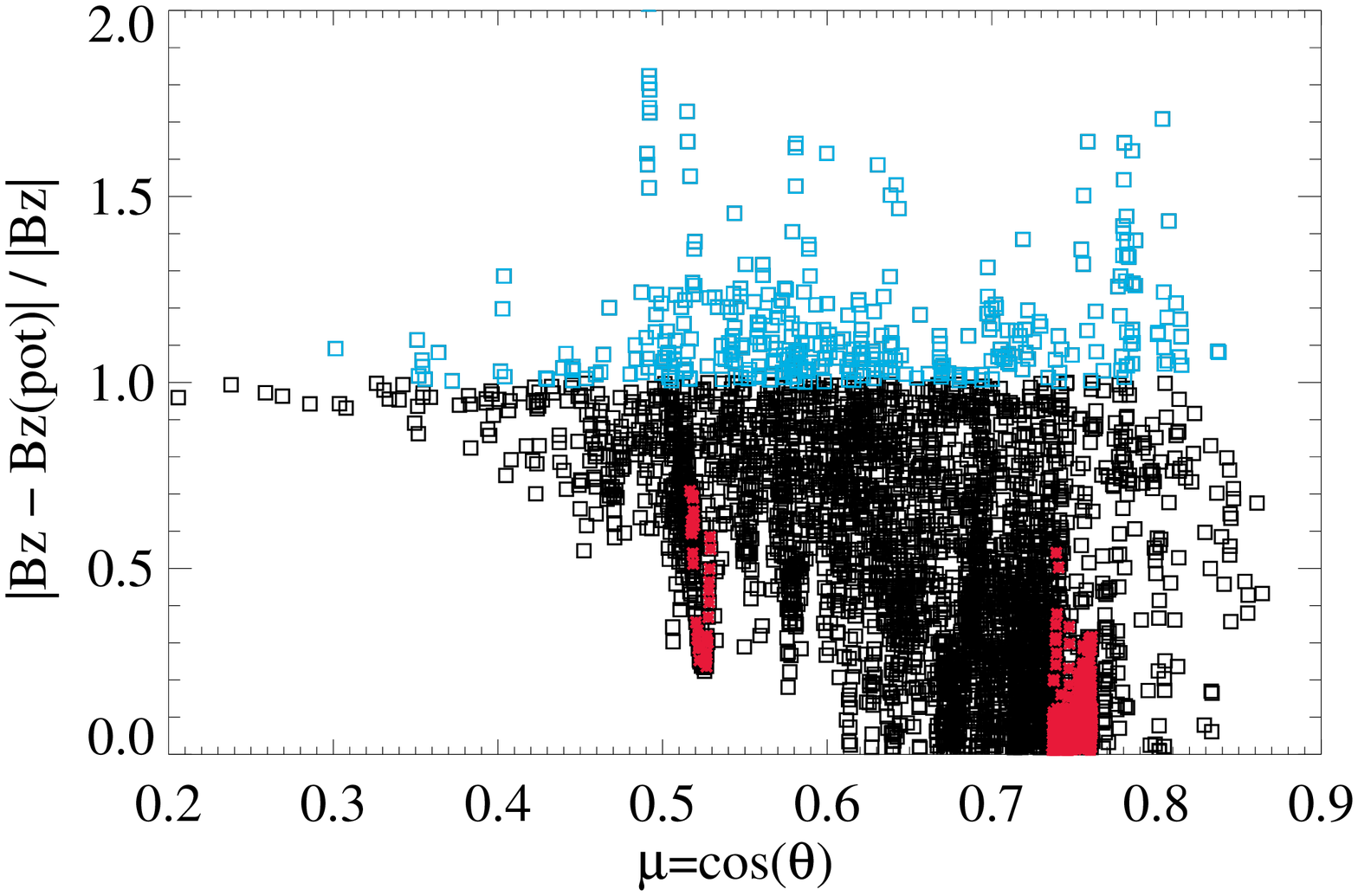}}
\caption{Same as Figure~\ref{fig:H3848diffs_norm} but each difference is normalized
by $|\Bz|$, to show the fractional change.  As such, those points with a sign
error (blue-square overplot) will all show a change of at least 100\%.}
\label{fig:H3848diffs_norm}
\end{figure}

To explore more where the different approaches fail, and how, an
analysis of the difference between the $\Br$ approximations and the
$\Bz$ component from the vector field observations is performed in
detail for one HMI Active Region Patch.  HARP~3848, observed by HMI
on 2014.03.15 at 15:48:00~TAI includes NOAA~AR~12005 and AR~12007
(Figure~\ref{fig:H3848context}), and was centered north-east of
disk center; it is one of the regions/days included in the larger
HARP dataset, and chosen because of its location, its relatively
simple main sunspot plus a second sunspot at a different $\mu$ value,
and that it includes a spread of plage over a fair range of $\mu$ as
well.  The differences between the observed radial component (in this
case all using planar approximations) $\Bz$ and two different $\Br$
approximations, from $\Brms$ and $\Bzppc$ are examined for points in
very restrictive local inclination ranges, as a function of structure
and observing angle.  Two representations are shown: the absolute
magnitude (Figure~\ref{fig:H3848diffs}) and the fractional difference
(Figure~\ref{fig:H3848diffs_norm}); those points which resulted in an
erroneous sign change (relative to the $\Bz$ boundary) are also indicated.
Only points which have a ``good'' disambiguation and have a signal/noise
ratio greater than 5.0 are included.

Summarizing both plots, for $\gamma < 10^\circ$, the $\Brms$ approximation
is systematically closer to $\Bz$ than the $\Bzppc$ results; this is
especially true in plage for the $\Brms$ model, which also shows no
sign-error points whereas there are a few sign-error points in the
$\Bzppc$ model.  For $20^\circ < \gamma < 30^\circ$, slightly inclined
fields, the bulk of the points are less different between the two
boundaries, however most striking is that the $\Brms$ approximation
is beginning to show significant differences in the sunspots, whereas
the $\Bzppc$ sunspot areas continue to display fairly small errors.
The latter do, however, show a greater number of plage points which
have introduced an erroneous polarity difference whereas the $\Brms$
shows these errors only at the more extreme $\mu$ values.  Examining only
points within $40^\circ<\gamma<50^\circ$ range -- significantly inclined
but not horizontal -- the errors in the spot become very large in $\Brms$,
but stay consistent and small in the $\Bzppc$ boundary.  More points in
the former are also of the incorrect sign, including many with quite large
field-strength differences.  The $\Bzppc$ boundary in fact performs better
for both plage and spots at these larger inclination angles than $\Brms$.

The appropriateness of a $\mu$-correction in the context of vertical
fields is shown thus to be true, but surprisingly limited to a very small
degree of deviation {\it away} from truly vertical.  By 20$^\circ$ from
vertical, the results are mixed.  For more inclined fields, the $\Brms$
is clearly problematic, particularly in sunspots (in part due to the
(unknown) inherent azimuthal angle), but also in inclined weak-field
areas.  One must note, however, that what are inferred to be weak-field
inclined points may be predominantly a product of noise in the vector
field data at these larger observing angles.  The $\Bzppc$ boundary
is demonstrably better for all inclinations within sunspots, and is
susceptible to polarity errors in weak-field areas at all inclinations
but at a somewhat more consistent, lower level.  In this particular case,
the percentage of plage-area pixels with incorrect sign (over all inherent
inclination angles) is lower for the $\Bzppc$ boundary than the $\Brms$
(6.1\% {\it vs.} 11\%, respectively), and the former appears to perform
quantitatively better for both plage and spot areas in this example.

\section{Conclusions}

A method is developed, based on earlier publications
\cite{Sakurai1982,Alissandrakis1981}, and tested here for its ability to
produce an estimate of the radial field distribution from line-of-sight
magnetic field observations of the solar photosphere.  Comparisons were
made between the line-of-sight component calculated from the vector-field
observations, the inferred radial-directed component from the same,
and different implementations of two approaches for estimating the
radial component from line-of-sight observations: one approach based
on the common ``$\mu$-correction'' and one based on using the radial
field component from a potential field constructed so as to match
the line-of-sight input.  The potential-field constructs impose, of
course, significant assumptions regarding the underlying structure.  The
``$\mu$-correction'' approach imposes a single much stronger assumption:
that the underlying field is always directed normal to the local surface.

We find that answering the question of which approach better recovers a target
quantity differs according to said target's underlying magnetic field
structure as would be inferred by the instrument at hand.  Structures
which abide by the radial-field assumption are well recovered by
``$\mu$-correction'' approaches.  This may, with caution, be extended to
structures whose inherent inclination is up to a few tens of degrees,
but with an uncertain worsening error as a function of increasing
observing angle.  Magnetic structures which would be observed to be
inherently more inclined by that instrument are generally poorly served by
``$\mu$-correction'' approaches.

Most active regions comprise a mix of structures, and as such making
general performance statements is dangerous.  While sunspot field
strengths are far better recovered using the potential-field constructs,
tests on polar or high-latitude areas that should be primarily plage
tell a mixed story: higher field strengths are recovered more reliably
using some form of $\mu$-correction, yet results also indicate that
a significant subset of the measurements are returned with the incorrect 
magnetic polarity.  That being said, the total magnetic
flux, an extensive quantity that often encompasses all structures within
an active region, can be better recovered by $\mu$-correction approaches
{\it if} the target is dominated by field which would expected to be inferred as
radial by the instrument involved.

$\Bl$ images of sunspots often have pronounced ``false'' magnetic 
polarity inversion lines; because the $\mu$-correction approaches
involve multiplying by a simple scaling factor, they cannot
relocate incorrect PILs and instead enhance incorrect-polarity field strengths. 
This is a particular problem when strong-gradient PILS need to 
be identified.  The potential-field approaches can mitigate
the false-PIL problem; the impact was demonstrated on a near-limb 
active region, but PIL displacement can occur at any location where
$\mu \neq 1.0$.

Of course it can be argued that the potential field is not appropriate
for magnetically complex active regions, and that linear or non-linear
extrapolations would perform even better.  Unfortunately, without crucial
additional constraints, there is no unique linear or nonlinear force-free
field solution to the $\Bl$ boundary, whereas the potential field provides
a unique solution.

The most general conclusions are first, that any correction improves
upon the naive $\Bl = \Br$ approach.  Second, the $\mu$-corrections
recover field strengths in areas inherently comprised of vertical
structures (as would be inferred by the instrument), but introduce
random errors whose magnitude can be surprisingly large given a
sunspot's proximity to disk center, and these corrections exacerbate
the influence of projection-induced sign errors.  Lastly, while the
potential-field reconstructions will introduce systematic errors,
generally underestimating field strengths and introducing new
polarity-sign errors in weaker and more radially-directed fields, it
recovers well both the radial-component field strengths in sunspots and
the locations of the magnetic polarity inversion lines.

\begin{acks}

The authors would like to acknowledge the helpful comments by the referee for this
manuscript.
The development of the potential-field reconstruction codes
and the preparation of this manuscript was supported by NASA through
contracts NNH12CH10C, 
NNH12CC03C, and  
grant NNX14AD45G. 
This material is based upon work supported by the National Science Foundation
under Grant No. 0454610.  Any opinions, findings, and conclusions or
recommendations expressed in this material are those of the author(s) and do
not necessarily reflect the views of the National Science Foundation.

\end{acks}

\appendix
\section{Method: Planar Approximation} 
\label{app:method_planar}

The line-of-sight component is observed on an image-coordinate planar grid. To
avoid having to interpolate to a regular heliographic grid, consider a hybrid
(non-orthogonal) coordinate system, $(\xi,\eta,z^h)$, consisting of the
transverse image components, $(\xi,\eta)$ and the heliographic normal
component, $z^h$. 

Following the convention of \inlinecite{garyhagyard90}, the coordinates $\xi$,
$\eta$ are defined in the $z^h=0$ plane in terms of heliographic coordinates by
\begin{eqnarray}
\left ( \begin{array}{c} \xi \\ \eta \end{array} \right ) &=& 
\left ( \begin{array}{cc} c_{11} & c_{12} \\ c_{21} & c_{22} \end{array} \right ) 
\left ( \begin{array}{c} x^h \\ y^h \end{array} \right ). 
\end{eqnarray} 
thus the new coordinate system is related to helioplanar coordinates by 
\begin{eqnarray}
\left ( \begin{array}{c} \xi \\ \eta \\ z^h \end{array} \right ) &=& 
\left ( \begin{array}{ccc} c_{11} & c_{12} & 0 \\ c_{21} & c_{22} & 0 \\ 
0 & 0 & 1 \end{array} \right ) 
\left ( \begin{array}{c} x^h \\ y^h \\ z^h \end{array} \right ). 
\end{eqnarray} 
Hence, the calculations are performed in the coordinate system $(\xi,\eta,z)$,
which is {\em not} the same as the image coordinates, except at $z=0$. As such,
there will be three coordinate systems under consideration, related as follows.
The heliographic and image coordinates are related by the standard transform
given in \inlinecite{garyhagyard90}: 
\begin{eqnarray}
\left ( \begin{array}{c} x^h \\ y^h \\ z^h \end{array} \right ) &=& 
\left ( \begin{array}{ccc} a_{11} & a_{12} & a_{13} \\ a_{21} & a_{22} & a_{23} \\ 
a_{31} & a_{32} & a_{33} \end{array} \right ) 
\left ( \begin{array}{c} x^i \\ y^i \\ z^i \end{array} \right ), 
\end{eqnarray} 
while the new coordinate system is related to the originals by 
\begin{eqnarray}
\left ( \begin{array}{c} \xi \\ \eta \\ z \end{array} \right ) &=& 
\left ( \begin{array}{ccc} c_{11} & c_{12} & 0 \\ c_{21} & c_{22} & 0 \\ 
0 & 0 & 1 \end{array} \right ) 
\left ( \begin{array}{c} x^h \\ y^h \\ z^h \end{array} \right ), 
\nonumber \\ 
&=& \left ( \begin{array}{ccc} c_{11} a_{11} + c_{12} a_{21} & c_{11} a_{12} + c_{12} a_{22} & c_{11} a_{13} + c_{12} a_{23} \\ 
c_{21} a_{11} + c_{22} a_{21} & c_{21} a_{12} + c_{22} a_{22} & c_{21} a_{13} + c_{22} a_{23} \\ 
a_{31} & a_{32} & a_{33} \end{array} \right ) 
\left ( \begin{array}{c} x^i \\ y^i \\ z^i \end{array} \right ). 
\end{eqnarray} 
Henceforth, the superscript on the heliographic components is dropped, but
retained on the image components. 

The volume of interest is restricted to $0<\xi<L_x$, $0<\eta<L_y$ and $z\ge0$.
Assuming that $c_{ij}$ is constant (that is, neglecting curvature across the
field of view), this transformation is linear and the solution to
Laplace's equation should still be of the form 
\begin{eqnarray}
\Phi(\xi,\eta,z) &=& \sum_{m,n} A_{mn} 
e^{[2 \pi i m \xi/L_x + 2 \pi i n \eta/L_y - \kappa_{mn} z]} + 
A_\xi \xi + A_\eta \eta + A_0 z, 
\end{eqnarray} 
with the value of $\kappa_{mn}$ determined by $\grad^2 \Phi=0$, namely 
\begin{eqnarray}
\grad^2 \Phi(\xi,\eta,z) &=& {\partial^2 \Phi \over \partial \xi^2} 
\bigg [\bigg ({d\xi \over dx} \bigg )^2 + 
\bigg ({d\xi \over dy} \bigg )^2 \bigg ] + 
2 {\partial^2 \Phi \over \partial \xi \partial \eta} 
\bigg [{d\xi \over dx} {d\eta \over dx} + 
{d\xi \over dy} {d\eta \over dy} \bigg ] 
\nonumber \\ 
&& + {\partial^2 \Phi \over \partial \eta^2} \bigg [\bigg ({d\eta \over dx} \bigg )^2 + 
\bigg ({d\eta \over dy} \bigg )^2 \bigg ] + {\partial^2 \Phi \over \partial z^2}
\nonumber \\ 
&=& \sum_{m,n} A_{mn} e^{[2 \pi i m \xi/L_x + 2 \pi i n \eta/L_y - \kappa_{mn} z]} 
\bigg \lbrace (c_{11}^2 + c_{12}^2) \bigg ({2 \pi i m \over L_x} \bigg)^2 
\nonumber \\ 
&& \qquad + 2 (c_{11} c_{21} + c_{12} c_{22}) \bigg ({2 \pi i m \over L_x} \bigg) 
\bigg ({2 \pi i n \over L_y} \bigg) 
\nonumber \\
&& \qquad + (c_{21}^2 + c_{22}^2) \bigg ({2 \pi i n \over L_y} \bigg)^2 + 
\kappa_{mn}^2 \bigg \rbrace 
\nonumber \\ 
\Rightarrow \ \kappa_{mn}^2 &=& (2 \pi)^2 \bigg [(c_{11}^2 + c_{12}^2) \bigg ({m \over L_x} \bigg )^2 + 
2 (c_{11} c_{21} + c_{12} c_{22}) \bigg ({m \over L_x} \bigg ) \bigg ({n \over L_y} \bigg ) 
\nonumber \\ 
&& \qquad + (c_{21}^2 + c_{22}^2) \bigg ({n \over L_y} \bigg )^2 \bigg ]
\end{eqnarray} 
and choose $\kappa_{mn}>0$ so the field decreases with height. Also choose
$A_\xi=A_\eta=0$, so the constant field is purely vertical. This is equivalent
to specifying the boundary condition at large heights. 

The line of sight component of the field is thus given by 
\begin{eqnarray}
B_l &=& {\partial \over \partial x_l} \bigg \lbrace \sum_{m,n} A_{mn} 
e^{2 \pi i m \xi/L_x + 2 \pi i n \eta/L_y - \kappa_{mn} z} + A_0 z \bigg \rbrace 
\nonumber \\ 
&=& a_{33} A_0 + \sum_{m,n} A_{mn} e^{2 \pi i m \xi/L_x + 2 \pi i n \eta/L_y - \kappa_{mn} z} 
\nonumber \\ 
&& \qquad \times \bigg [{2 \pi i m \over L_x} (c_{11} a_{13} + c_{12} a_{23}) + 
{2 \pi i n \over L_y} (c_{21} a_{13} + c_{22} a_{23}) - \kappa_{mn} a_{33} \bigg ]. 
\end{eqnarray} 
Solve for the coefficients $A_{mn}$ by taking the Fourier transform of the 
line of sight component of the field at the surface
\begin{eqnarray}
{\tt FFT}(B^l) &=& \int_0^{L_x} d\xi \, \int_0^{L_y} d\eta \, B_l(\xi,\eta,0) 
e^{- 2 \pi i j \xi/L_x - 2 \pi i k \eta/L_y} 
\nonumber \\ 
&=& \sum_{m,n} A_{mn} \bigg [{2 \pi i m \over L_x} (c_{11} a_{13} 
+ c_{12} a_{23}) + {2 \pi i n \over L_y} (c_{21} a_{13} + c_{22} a_{23}) 
- \kappa_{mn} a_{33} \bigg ]
\nonumber \\ 
&& \qquad \times \int_0^{L_x} d\xi \int_0^{L_y} d\eta 
e^{2 \pi i (m - j) \xi/L_x + 2 \pi i (n - k) \eta/L_y} 
\nonumber \\ 
&& \quad + A_0 {\partial z \over \partial x_l} 
\int_0^{L_x} d\xi \, \int_0^{L_y} d\eta \, e^{- 2 \pi i j \xi/L_x - 2 \pi i k \eta/L_y} 
\nonumber \\ 
&=& {L_x} {L_y} A_{jk} \bigg [{2 \pi i j \over L_x} (c_{11} a_{13} + c_{12} a_{23}) + 
{2 \pi i k \over L_y} (c_{21} a_{13} + c_{22} a_{23}) - \kappa_{jk} a_{33} \bigg ]
\nonumber \\ 
&& \quad + L_x L_y a_{33} A_0 \delta_{0j} \delta_{0k}. 
\end{eqnarray}
Knowing the coefficients, the vertical component of the field is given by 
\begin{eqnarray}
B_z(\xi,\eta,0) &=& {\partial \over \partial z} \bigg \lbrace \sum_{m,n} A_{mn} 
e^{2 \pi i m \xi/L_x + 2 \pi i n \eta/L_y - \kappa_{mn} z} + A_0 z \bigg \rbrace 
\bigg \vert_{z=0} 
\nonumber \\ 
&=& A_0 - \sum_{m,n} \kappa_{mn} A_{mn} e^{2 \pi i m \xi/L_x + 2 \pi i n \eta/L_y}. 
\end{eqnarray} 

\section{Method: Spherical Case}
\label{app:method_spherical}

Following the derivation given in \inlinecite{altschulernewkirk69}, but also see 
\inlinecite{Bogdan86}, the
potential field in a semi-infinite volume $r \ge R$ can be written in terms of
a scalar potential given by
\begin{eqnarray} 
\Psi &=& R \sum_{n=1}^\infty \sum_{m=0}^n \bigg ({R \over r} \bigg )^{n+1} 
(g_n^m \cos m \phi + h_n^m \sin m \phi) P_n^m(\mu),
\end{eqnarray} 
where $\mu=\cos \theta$, in terms of which the heliographic components of the
field are given by 
\begin{eqnarray} 
B_r &=& - {\partial \Psi \over \partial r} 
\nonumber \\ 
&=& \sum_{n=1}^\infty \sum_{m=0}^n (n + 1) \bigg ({R \over r} \bigg )^{n+2} 
(g_n^m \cos m \phi + h_n^m \sin m \phi) P_n^m(\mu),
\\ 
B_\theta &=& - {1 \over r} {\partial \Psi \over \partial \theta} 
\nonumber \\ 
&=& {1 \over \sin \theta} \sum_{n=1}^\infty \sum_{m=0}^n 
\bigg ({R \over r} \bigg )^{n+2} (g_n^m \cos m \phi + h_n^m \sin m \phi) 
\nonumber \\
&& \qquad \times \bigg [(n + 1) \mu P_n^m(\mu) - (n - m + 1) P_{n+1}^m(\mu) 
\bigg ] 
\\ 
B_\phi &=& - {1 \over r \sin \theta} {\partial \Psi \over \partial \phi} 
\nonumber \\ 
&=& {1 \over \sin \theta} \sum_{n=1}^\infty \sum_{m=0}^n m 
\bigg ({R \over r} \bigg )^{n+2} (g_n^m \sin m \phi - h_n^m \cos m \phi) 
P_n^m(\mu).
\end{eqnarray} 

Following \inlinecite{Rudenko2001a}, define the coordinate system such that the line
of sight direction corresponds to the polar axis of the expansion.  With this
choice, the line of sight component of the field is given by
\begin{eqnarray}\label{eqn:Blos} 
B_l &=& B_r \cos \theta - B_\theta \sin \theta. 
\nonumber \\ 
&=& \sum_{n=1}^\infty \sum_{m=0}^n (n - m + 1) \bigg ({R \over r} \bigg )^{n+2} 
P_{n+1}^m(\mu) \bigg [g_n^m \cos m \phi + h_n^m \sin m \phi \bigg ]. 
\end{eqnarray} 
To determine the coefficients in the expansion, first multiple both sides of 
equation~\ref{eqn:Blos} by $\cos m \theta$, and integrate over the surface of
the sphere of radius $R$:
\begin{eqnarray} 
&& \int_0^\pi \sin \theta d\theta \int_0^{2\pi} d\phi \, \cos m \phi 
P_{n+1}^m(\mu) B_l(R,\theta,\phi)  
\nonumber \\ 
&=& \sum_{n'=1}^\infty \sum_{m'=0}^{n'} \int_0^\pi \sin \theta d\theta 
\int_0^{2\pi} d\phi \, \cos m \phi (n' - m' + 1) P_{n+1}^m(\mu) 
P_{n'+1}^{m'}(\mu) 
\nonumber \\ 
&& \quad \times \bigg [g_{n'}^{m'} \cos m' \phi + h_{n'}^{m'} \sin m' \phi 
\bigg ] 
\nonumber \\ 
&=& {4 \pi g_n^m \over 2 n + 3} {(n + m + 1)! \over (n - m)!}, 
\end{eqnarray} 
which determines $g_n^m$.  Next, multiply both sides of equation~\ref{eqn:Blos}
by $\sin m \theta$, and again integrate over the surface of the sphere of
radius $R$:
\begin{eqnarray} 
&& \int_0^\pi \sin \theta d\theta \int_0^{2\pi} d\phi \, \sin m \phi 
P_{n+1}^m(\mu) B_l(R,\theta,\phi)  
\nonumber \\ 
&=& \sum_{n'=1}^\infty \sum_{m'=0}^{n'} \int_0^\pi \sin \theta d\theta 
\int_0^{2\pi} d\phi \, \sin m \phi (n' - m' + 1) P_{n+1}^m(\mu) 
P_{n'+1}^{m'}(\mu) 
\nonumber \\ 
&& \quad \times \bigg [g_{n'}^{m'} \cos m' \phi + h_{n'}^{m'} \sin m' \phi 
\bigg ] 
\nonumber \\ 
&=& {4 \pi h_n^m \over 2 n + 3} {(n + m + 1)! \over (n - m)!} 
\end{eqnarray} 
to determine $h_n^m$.

Because observations are only available for the near side of the Sun, to
actually implement this, it is necessary to make an assumption about the far
side of the Sun.  In order to ensure a zero monopole moment, it is convenient
to make the field anti-symmetric in some form. One convenient way to do this is
to let $B_l(R,\pi-\theta,\phi)=B_l(R,\theta,\phi)$, where the front side of the
Sun is assumed to lie in the range $0<\theta<\pi/2$.  Using the fact that the
associated Legendre functions have the property
\begin{eqnarray}
P_n^m(-x) &=& (-1)^{(n+m)} P_n^m(x),
\end{eqnarray}
the expressions for the coefficients become
\begin{eqnarray} 
g_n^m &=& {(2 n + 3) (n - m)! \over 4 \pi (n + m + 1)!} \int_0^{2\pi} d\phi \, 
\cos m \phi 
\nonumber \\
&& \qquad \times \bigg \lbrace \int_{-1}^0 d\mu \, P_{n+1}^m(\mu) 
+ \int_0^1 d\mu \, P_{n+1}^m(\mu) \bigg \rbrace B_l(R,\mu,\phi) 
\nonumber \\ 
&=& {(2 n + 3) (n - m)! [1 + (-1)^{n+m+1}] \over 4 \pi (n + m + 1)!} 
\nonumber \\ 
&& \qquad \times \int_0^{2\pi} d\phi \, \cos m \phi \int_0^1 d\mu \, 
P_{n+1}^m(\mu) B_l(R,\mu,\phi) 
\end{eqnarray} 
and similarly 
\begin{eqnarray} 
h_n^m &=& {(2 n + 3) (n - m)! [1 + (-1)^{n+m+1}] \over 4 \pi (n + m + 1)!} 
\nonumber \\
&& \qquad \times \int_0^{2\pi} d\phi \, \sin m \phi \int_0^1 d\mu \, 
P_{n+1}^m(\mu) B_l(R,\mu,\phi). 
\end{eqnarray} 
Note that the terms with $n+m$ even have $g_n^m=h_n^m=0$. This is a consequence
of the boundary condition imposed for the far side of the Sun, which
effectively reduces the number of independent terms by a factor of two. 


\end{article}
\end{document}